\documentclass[a4paper,fleqn,usenatbib, useAMS]{mnras}
\usepackage{graphicx}
\usepackage{multicol}
\usepackage{enumerate}
\usepackage{amsmath}
\usepackage{pdflscape}
\usepackage[utf8]{inputenc}
\usepackage{journal_names}
\usepackage{placeins}
\usepackage{amssymb}
\usepackage{longtable}
\usepackage[flushleft]{threeparttable}
\def\farcs{\hbox{$.\!\!^{\prime\prime}$}}

\def\hetwo{He\,\textsc{ii} }
\def\heone{He\,\textsc{i} }
\newcommand{\package}[1]{\texttt{#1}}
\title[Low ionisation emission lines in the TDE candidate AT~2018fyk] {Evidence for rapid disk formation and reprocessing in the X-ray bright tidal disruption event candidate AT~2018fyk}
\author[Wevers, T. et al.]{T. Wevers$^{1}$\thanks{Email: tw@ast.cam.ac.uk}, D.\,R. Pasham$^{2}$, S. van Velzen$^{3,4}$, G. Leloudas$^{5}$, S. Schulze$^{6}$,  \newauthor J.\,C.\,A. Miller-Jones$^{7}$, P.\,G. Jonker$^{8,9}$, M. Gromadzki$^{10}$, E. Kankare$^{11}$, S.\,T. Hodgkin$^1$,\newauthor \L{}. Wyrzykowski$^{10}$, Z. Kostrzewa-Rutkowska$^{8}$, S. Moran$^{11,12}$, M. Berton$^{13,14}$, \newauthor  K. Maguire$^{15,16}$, F. Onori$^{17}$, S. Mattila$^{11}$ and M. Nicholl$^{18,19}$\\\\
$^{1}$Institute of Astronomy, University of Cambridge, Madingley Road, Cambridge CB3 0HA, United Kingdom \\
$^{2}$MIT Kavli Institute for Astrophysics and Space Research, Cambridge, MA 02139, USA \\
$^{3}$Department of Astronomy, University of Maryland, College Park, MD 20742 \\
$^{4}$Center for Cosmology and Particle Physics, New York University, NY 10003 \\
$^{5}$DTU Space, National Space Institute, Technical University of Denmark, Elektrovej 327, 2800 Kgs.  Lyngby, Denmark \\
$^{6}$Department of Particle Physics and Astrophysics, Weizmann Institute of Science, Rehovot 7610001, Israel \\
$^{7}$ICRAR – Curtin University, GPO Box U1987, Perth, WA 6845, Australia \\
$^{8}$SRON, Netherlands Institute for Space Research, Sorbonnelaan 2, 3584 CA, Utrecht, The Netherlands \\
$^{9}$Department of Astrophysics/IMAPP, Radboud University, P.O. Box 9010, 6500 GL Nijmegen, The Netherlands \\
$^{10}$Warsaw University Astronomical Observatory, Al. Ujazdowskie 4, 00-478 Warszawa, Poland \\
$^{11}$Tuorla Observatory, Department of Physics and Astronomy, University of Turku, Väisäläntie 20, FI-21500 Piikkiö, Finland \\
$^{12}$Nordic Optical Telescope, Apartado 474, E-38700 Santa Cruz de La Palma, Spain \\
$^{13}$Finnish Centre for Astronomy with ESO (FINCA), University of Turku, Quantum, Vesilinnantie 5, FI-20014, University of Turku, Finland \\
$^{14}$Aalto University Mets{\"a}hovi Radio Observatory, Mets{\"a}hovintie 114, FI-02540 Kylm{\"a}l{\"a}, Finland \\
$^{15}$Astrophysics Research Centre, School of Mathematics and Physics, Queens University Belfast, Belfast BT7 1NN, UK \\
$^{16}$School of Physics, Trinity College Dublin, Dublin 2, Ireland \\
$^{17}$Istituto di Astrofisica e Planetologia Spaziali (INAF), via del Fosso del Cavaliere 100, Roma, I-00133, Italy \\
$^{18}$Institute for Astronomy, University of Edinburgh, Royal Observatory, Blackford Hill, EH9 3HJ, UK \\
$^{19}$Birmingham Institute for Gravitational Wave Astronomy and School of Physics and Astronomy, University of Birmingham,\\ Birmingham B15 2TT, UK }
\begin{document}
\pagerange{\pageref{firstpage}--\pageref{lastpage}} \pubyear{2019}
\maketitle
\label{firstpage}
\begin{abstract}
We present optical spectroscopic and {\it Swift} UVOT/XRT observations of the X-ray and UV/optical bright tidal disruption event (TDE) candidate AT~2018fyk/ASASSN--18ul discovered by ASAS--SN. The {\it Swift} lightcurve is atypical for a TDE, entering a plateau after $\sim$40 days of decline from peak. After 80 days the UV/optical lightcurve breaks again to decline further, while the X-ray emission becomes brighter and harder. In addition to broad H, He and potentially O/Fe lines, narrow emission lines emerge in the optical spectra during the plateau phase. We identify both high ionisation (O\,\textsc{iii}) and low ionisation (Fe\,\textsc{ii}) lines, which are visible for $\sim$45 days. We similarly identify Fe\,\textsc{ii} lines in optical spectra of ASASSN--15oi 330 d after discovery, indicating that a class of Fe-rich TDEs exists. The spectral similarity between AT~2018fyk, narrow-line Seyfert 1 galaxies and some extreme coronal line emitters suggests that TDEs are capable of creating similar physical conditions in the nuclei of galaxies. The Fe\,\textsc{ii} lines can be associated with the formation of a compact accretion disk, as the emergence of low ionisation emission lines requires optically thick, high density gas. Taken together with the plateau in X-ray and UV/optical luminosity this indicates that emission from the central source is efficiently reprocessed into UV/optical wavelengths. Such a two-component lightcurve is very similar to that seen in the TDE candidate ASASSN--15lh, and is a natural consequence of a relativistic orbital pericenter. 
\end{abstract}
\begin{keywords}
accretion, accretion disks -- galaxies: nuclei -- black hole physics -- ultraviolet: galaxies -- X-rays: galaxies
\end{keywords}

\section{Introduction}
Passing within the tidal radius of the supermassive black hole (SMBH) in the centre of a galaxy can lead to a star's demise \citep{Hills1975,Rees1988, Phinney1989}. Such cataclysmic events, called tidal disruption events (TDEs), resemble panchromatic cosmic fireworks, with bright emission at wavelengths ranging from radio \citep{vanvelzen2015,Alexander2016}, IR \citep{vanvelzen2016b,jiang2016, Matilla2018}, optical and UV \citep{Gezari2008,vanvelzen2011,Arcavi2014,Holoien201614li,Wyrzykowski2017} as well as X-rays \citep{Komossa1999, Greiner2000} and even $\gamma$ rays \citep{Bloom2011,Cenko2012}. The duration and brightness of such flares depends on the complex dynamics of material in the presence of strong gravitational fields \citep{Guillochon2015,Metzger2016}. Wide-field surveys such as the Roentgen Satellite (ROSAT) and the X-ray Multi-Mirror telescope (XMM; \citealt{Jansen2001}) in X-rays and the Galaxy and Evolution Explorer (GALEX), Sloan Digital Sky Survey (SDSS; \citealt{Stoughton2002}), the (intermediate) Palomar Transient Factory (PTF; \citealt{Law2009}) and the All Sky Automated Supernova (ASASSN; \citealt{asassn}) surveys in the UV/optical have led to the discovery and characterisation, first in archival data and later in near real-time, of a few dozen TDEs and even more TDE candidates. 

Sparse (or non-existent) temporal data coverage of UV/optical selected TDEs at X-ray wavelengths (and vice-versa) inhibit the multi-wavelength characterisation and subsequently the detailed study of the energetics and dynamics at play. This sparse coverage is the result of a variety of factors, such as the difficulty to perform image subtraction in galactic nuclei, the need for fast and systematic spectroscopic follow-up of nuclear transients and the limited availability of multi-wavelength monitoring. Coordinated efforts in recent years have led to significant progress in this respect, and most spectroscopically confirmed TDEs are now observed with the {\it Swift} X-ray observatory, made possible due to its flexible scheduling system.

Nevertheless, disentangling the dominant emission mechanisms remains a challenge. The thermal soft X-ray emission is thought to originate from a compact accretion disk (e.g. \citealt{Komossa1999,Auchettl2017}) while luminous hard X-ray emission finds it origin in a relativistic jet \citep{Bloom2011, Cenko2012}. For the UV/optical emission, however, a clear picture has not yet emerged. Shocks due to stream-stream collisions \citep{Piran2015, Shiokawa2015} or reprocessing of accretion power in either static \citep{Loeb1997, Guillochon2014, Roth2016} or outflowing material (e.g. \citealt{Strubbe2009,Metzger2016,Roth2018}) have all been proposed to explain the observations. 
\citet{Dai2018} proposed a model that can explain both the X-ray and UV/optical observations by suggesting a geometry similar to the active galactic nucleus (AGN) unification model (see also \citealt{Metzger2016}), where an optically thick structure in the disk orbital plane or an optically thick super-Eddington disk wind obscures the X-ray emission for certain viewing angles. The presence of Bowen fluorescence lines, which require an X-ray powering source, in several TDEs with X-ray non-detections \citep{Leloudas2019}, support this scenario.

In terms of their optical spectra, TDEs typically show broad (10--20 $\times10^3$ km s$^{-1}$) H and/or He lines \citep{Arcavi2014}, although it is unclear what determines whether a TDE is H-rich, He-rich or shows both features. Furthermore, while some TDEs show only broad He\,\textsc{ii} emission, the sudden appearance or disappearance of other lines such as He\,\textsc{i} has been observed \citep{Holoien201615oi}. One feature in particular is observed in many TDEs: the broad He\,\textsc{ii} line appears to have an asymmetric shoulder in its blue wing. Moreover, it is often observed to be significantly blueshifted (when fit with a Gaussian line profile), whereas other broad Balmer lines, when present, do not show a similar blueshift. 
While asymmetric Balmer emission line profiles can be modelled using an elliptical accretion disk model \citep{Liu2017,Cao2018,Holoien2018} or alternatively a spherically expanding medium \citep{Roth2018, Hung2019}, it does not appear to adequately explain the He\,\textsc{ii} line morphology. \citet{Leloudas2019} suggest instead that the asymmetry in the line is due to Bowen fluorescence lines, but this cannot explain {\it all} cases (e.g. ASASSN--15oi).

\citet{Leloudas2016} were the first to claim that two emission mechanisms were observed in a TDE candidate, namely in the double-peaked lightcurve of ASASSN--15lh. Although the debate as to the nature of this peculiar transient event is still ongoing \citep{Dong2016, Godoy2017,Margutti2017}, one explanation focused on the TDE interpretation. \citet{Leloudas2016} claim that the double-peaked lightcurve can be explained in terms of the fallback and viscous timescales around a very massive ($\geq$10$^8$ M$_{\odot}$) SMBH. In this case the orbital pericenter of the disrupted star is relativistic, making disk formation very efficient. This can lead to two distinct maxima in the lightcurve. In fact, \citet{vanvelzen2019} recently demonstrated that a two-phase structure appears to be common for all TDEs, but often the second, more shallow phase is observed a few years after peak. Alternatively, \citet{Margutti2017} invoke a model where a sudden change in the ejecta opacity due to an underlying source of ionising radiation leads to a double-humped lightcurve. We will show that the lightcurve of AT~2018fyk shows a similarly double-humped profile to ASASSN--15lh. We propose that the relatively massive black hole ($\sim$2$\times$10$^7$ M$_{\rm BH}$) for AT2018fyk similarly leads to a relativistic pericenter, speeding up the disk formation process and explaining the similarities. 

In this work we present our observations of a new tidal disruption event candidate, AT~2018fyk/ASASSN--18ul, discovered by the All Sky Automated Survey for SuperNovae (ASAS--SN; \cite{Shappee2014}). We analyse {\it Swift} UVOT and XRT data together with optical low resolution spectroscopic observations covering the first 120 days of its evolution. While both the lightcurve and spectra show features peculiar to known TDEs, in particular a secondary maximum in the UVOT bands and the simultaneous emergence of narrow emission lines (in addition to broad H and He lines), we show that these properties can be explained by the reprocessing of (part of the) X-ray emission into UV/optical photons. While the lightcurve is similar to ASASSN--15lh, this is the first time that unambiguous evidence for reprocessing is found in the optical spectra of TDEs. This shows that the dynamics of the disruption can leave clear imprints on the lightcurves. Furthermore, the spectral signatures of reprocessing are strongest during the second maximum in the lightcurve. This suggests that the X-ray source turned on almost contemporaneously with the initial UV/optical peak, in line with a rapid accretion disk formation scenario.

In Section \ref{sec:observations}, we present X-ray, UV/optical and radio observations and describe the data reduction process. We present the spectroscopic and lightcurve analysis and results in Section \ref{sec:analysis}, while discussing the implications in Section \ref{sec:discussion}. We summarise our main findings in Section \ref{sec:summary}.

\begin{figure*}
\includegraphics[width=0.65\textwidth, keepaspectratio]{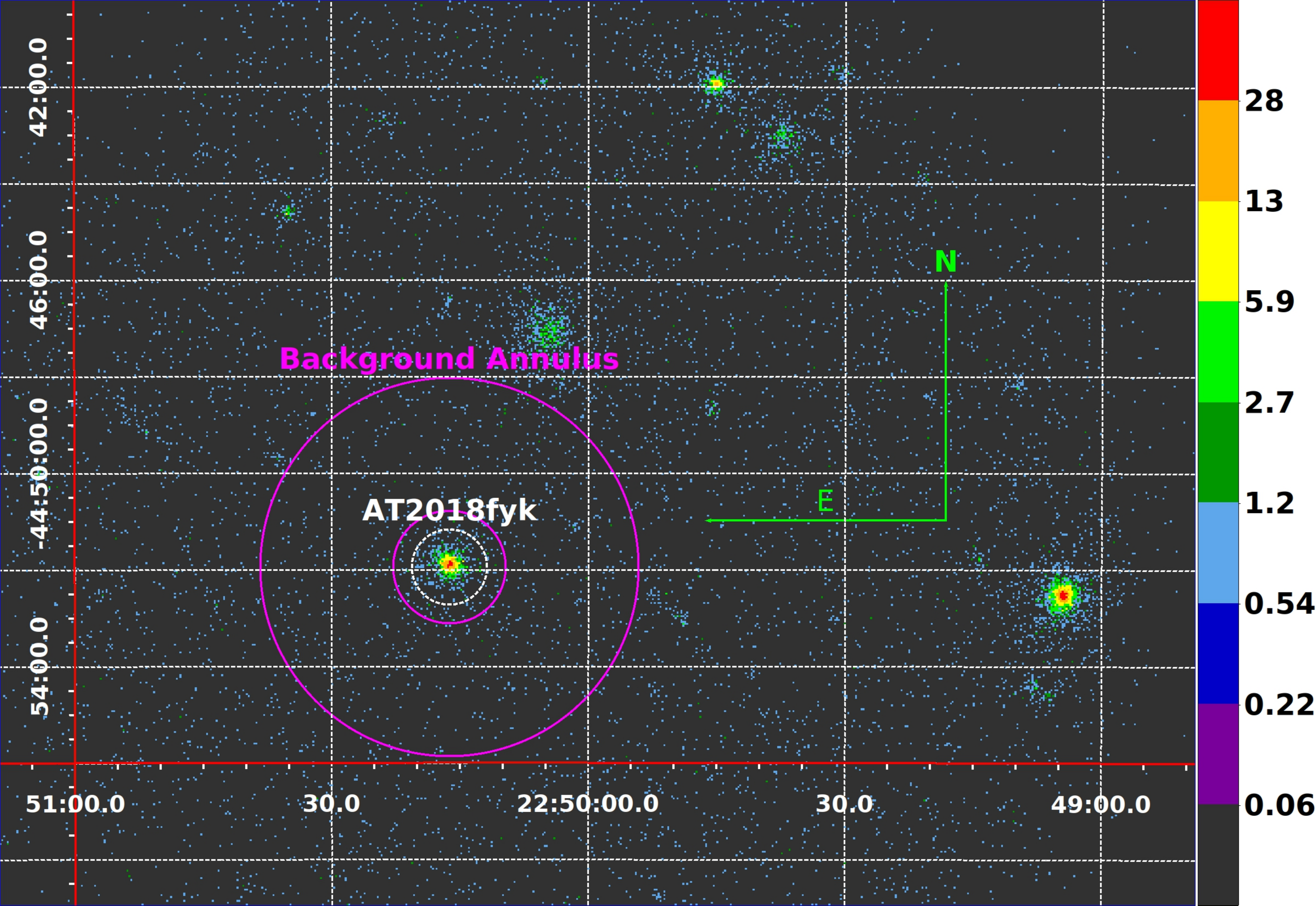}
\caption{X-ray (0.3-8.0 keV) image of AT~2018fyk's {\it Swift}/XRT field of view. The source extraction region is indicated by a white dashed circle with a radius of 47$^{\prime\prime}$. The background count rates from each XRT exposure were estimated within an annular region (magenta) with inner and outer radii of 70$^{\prime\prime}$ and 235$^{\prime\prime}$, respectively. The green arrows are each 300$^{\prime\prime}$.}
\label{fig:fov}
\end{figure*}

\section{Observations and data reduction}
\label{sec:observations}
The transient AT~2018fyk/ASASSN--18ul was discovered near the center of the galaxy LCRS B224721.6-450748 (estimated offset of 0.85 arcsec from the nucleus) by the ASAS--SN survey on 2018 September 8 (MJD 58\,369.23). The estimated transient brightness was $g$=17.8 mag, with a non-detection reported ($g\geq$17.4 mag) on 2018 August 29. A classification spectrum was taken as part of the extended Public ESO Spectroscopic Survey for Transient Objects (ePESSTO; \citealt{Smartt2015}) on 2018 September 15, revealing a blue featureless continuum superposed with several broad emission lines, suggesting that the transient was likely a TDE \citep{wevers:fykopticalspectrum}.

No high spatial resolution archival imaging is available to constrain the position of the transient with respect to the host galaxy centre of light. 
Fortunately, Gaia Science Alerts (GSA; \citealt{Hodgkin2014}) also detected the transient (aka Gaia18cyc) at the position ($\alpha$,$\delta$) = (22:50:16.1, --44:51:53.5) on 2018 October 10, with an estimated astrometric accuracy of $\sim$100 mas\footnote{This is due to the fact that GSA uses the initial data treatment astrometric solution \citep{Fabricius2016}. In the future, the implementation of an improved astrometric solution could improve this to mas precision.}. The host galaxy is part of the Gaia Data Release 2 (GDR2) catalogue \citep{Brown2016,Prusti2018}, and its position is reported as ($\alpha$,$\delta$) = (22:50:16.093, --44:51:53.499) with formal uncertainties of 1.1 and 1.5 mas in right ascension and declination, respectively \citep{Lindegren2018}. We note that the GDR2 astrometric$\_$excess$\_$noise parameter is 11 mas, which indicates that the formal errors are likely underestimated (as expected for an extended source, \citealt{Lindegren2018}). The offset between the transient and host galaxy positions is 15 mas. 

\citet{Kostrzewa2018} have shown that the mean offset in the Gaia data of SDSS galaxies is $\sim$100 mas, consistent with the mean offset of SDSS galaxies and their GDR2 counterparts. Additionally, we can try to estimate a potential systematic offset between Gaia transients and their GDR2 counterparts. To quantify such an offset, we crossmatch the $\sim$7000 published Gaia alerts with GDR2 within a search radius of 0.25 arcsec. The offset distribution (angular distance on the sky) is well described by a Rayleigh function, as expected if the uncertainties in right ascension and declination follow a normal distribution. The distance distribution has a median of 62 mas and standard deviation of 40 mas. This represents the potential systematic offset between the coordinate systems and is fully consistent with the 100 mas transient positional uncertainties, indicating that both coordinate systems are properly aligned.

In conclusion, we find an offset between the transient and host galaxy position of 15$\pm$100 mas, which corresponds to 17$\pm$120 pc at the host redshift. This illustrates the power of Gaia for identifying nuclear transients (see also \citealt{Kostrzewa2018} for a detailed investigation), as it firmly constrains AT~2018fyk to the nucleus of the galaxy. 

\subsection{Host galaxy spectral energy distribution}
We determine the host galaxy redshift from the spectra, which show strong Ca\,\textsc{ii} H+K absorption lines, and find z=0.059. This corresponds to a luminosity distance of approximately 275.1 Mpc, assuming a $\Lambda$CDM cosmology with $H_0=67.11~{\rm km\,s}^{-1}\,{\rm Mpc}^{-1}$, $\Omega_m=0.32$ and $\Omega_\Lambda=0.68$ \citep{Planck2014a}.
No narrow emission lines from the host galaxy are evident, indicating that the event occurred in a quiescent galaxy. We observe H$\alpha$ and H$\beta$ in absorption, indicating no ongoing star formation. The lack of significant H$\delta$ absorption suggests that the galaxy does not belong to the E+A galaxy class \citep{Dressler1983} in which TDEs have been known to be overrepresented \citep{Arcavi2014,French2016}. We identify strong absorption lines at $\lambda$4303 (G-band), $\lambda$5172 (Mg\,\textsc{i}\,$b$, which indicates an old stellar population), $\lambda$5284 (Fe\,\textsc{ii}) and the Na\,\textsc{i} D doublet at 5890+5895 \AA. Finally, the AllWISE color $W1$--$W2$=0.04 \citep{Cutri2014} further indicates that the black hole is most likely inactive (e.g. \citealt{Wu2012,Stern2012}).

To measure the galaxy mass and star formation rate (SFR), we model the spectral energy distribution (SED; see Table \ref{tab:hostgalphot}) with the software package \package{LePhare} version 2.2 \citep{Arnouts1999a,Ilbert2006a}\footnote{\href{http://www.cfht.hawaii.edu/~arnouts/LEPHARE/lephare.html}{http://www.cfht.hawaii.edu/\~{}arnouts/LEPHARE/lephare.html}}. This also allows us to synthesise the host galaxy brightness in the {\it Swift} bands, which we use to subtract the host galaxy contribution from the TDE lightcurves. We generate $3.9\times10^6$ templates based on the \citet{Bruzual2003a} stellar population synthesis models with the Chabrier initial mass function \citep[IMF;][]{Chabrier2003a}. The star formation history (SFH) is approximated by a declining exponential function of the form e$^{(-t/ \tau)}$, where $t$ is the age of the stellar population and $\tau$ the e-folding time-scale of the SFH (varied in nine steps between 0.1 and 30 Gyr). These templates are attenuated with the Calzetti attenuation curve \citep[varied in 22 steps from $E(B-V)=0$ to 1 mag;][]{Calzetti2000a}. \package{LePhare} accounts for the contribution from the diffuse gas (e.g. \ion{H}{ii} regions) following the relation between SFR and the line fluxes presented in \citet{Kennicutt1998a}. 

From the best fit template spectrum, we derive a host galaxy stellar mass of log(M$_{\star}$/M$_{\odot}$)=10.2$^{+0.5}_{-0.2}$, and a SFR and intrinsic $E(B-V)$ consistent with 0. Using an empirical bulge-to-total (B/T) ratio \citep{Stone2018} of 0.47 (very similar to the ratio of the PSF to Petrosian $g$-band flux of 0.57) for this galaxy mass, we find a SMBH mass of 2$^{+3}_{-1.2} \times 10^{7}$ M$_{\odot}$ using the M$_{\rm BH}$--M$_{\rm bulge}$ relation \citep{Haring2004}. We synthesise photometry in the {\it Swift} UVOT filters, which can be found in Table \ref{tab:hostgalphot}, to perform the host subtraction. 

\begin{table}
\centering
\caption{Host galaxy photometry, both observed (above the line) and synthesised in the {\it Swift} UVOT bands (below the line). The synthetic {\it Swift} photometry is used for host galaxy subtraction of the lightcurves.}
\begin{tabular}{cc}\hline
Filter & AB mag \\\hline\hline
GALEX NUV & 21.91 $\pm$ 0.4\\
SkyMapper $g$ & 17.07$\pm$0.05\\
SkyMapper $r$ & 16.51$\pm$0.14\\
SkyMapper $i$ & 15.98$\pm$0.04\\
SkyMapper $z$ & 15.71$\pm$0.18\\
WISE $W1$ & 16.27$\pm$0.03\\
WISE $W2$ & 16.87$\pm$0.03\\\hline
Swift $UVW2$ & 22.3\\
Swift $UVM2$ & 21.9\\
Swift $UVW1$ & 20.8\\
Swift $U$ & 18.7\\
Swift $B$ & 17.4\\
Swift $V$ & 16.5\\
\hline
\end{tabular}
\label{tab:hostgalphot}
\end{table}

\subsection{{\it Swift} X-ray and UV/optical observations}
{\it Swift}'s \citep{swift:gehrels} UltraViolet/Optical Telescope (UVOT; \citealt{swift:uvot}) and the X-Ray Telescope (XRT; \citealt{swift:xrt}) started monitoring AT~2018fyk on MJD 58\,383.7, approximately 8 days after the classification spectrum was taken and 14 days after the reported discovery \citep{brima:fykdiscoveryatel} by the ASAS--SN survey. Between 2018 September 22 and 2019 January 8, 52 monitoring observations were made with an  average observing cadence varying between 2 and 4 days. {\it Swift} could not observe the source after 2019 January 8 due to Sun pointing constraints. We removed two observations (obsIDs: 00010883004 and 00010883038) from further analysis as they had limited XRT exposure ($\sim$ 10 s) and lacked UVOT data. Figure \ref{fig:fov} shows an X-ray image of AT~2018fyk's field of view as observed with {\it Swift}/XRT.

The XRT observations were all performed in photon counting (PC) mode, and were reduced using the latest version of the {\it Swift} \textsc{xrtpipeline} provided as part of {\tt Heasoft} 6.25 analysis package. Source counts were extracted using a circular aperture with a radius of 47$^{\prime\prime}$, and corrected for the background contribution using an annulus with an inner and outer radius of 70$^{\prime\prime}$ and 250$^{\prime\prime}$, respectively. Count rates are converted to an unabsorbed 0.3\,--\,8 keV flux using a conversion factor of 4.4\,$\times$\,10$^{-11}$, derived from the average count rate and flux in the stacked X-ray observations, and assuming a Galactic n$_{\rm H}$ column of 0.012\,$\times$\,10$^{20}$ cm$^{-2}$.

We note that no source is detected in archival ROSAT observations down to a limit of $\sim$5$\times$10$^{-4}$ cts s$^{-1}$ \citep{Boller2016}. Using the webPIMMS tool\footnote{https://heasarc.gsfc.nasa.gov/cgi-bin/Tools/w3pimms/w3pimms.pl}, this corresponds to a flux limit of 5$\times$10$^{-15}$ erg cm$^{-2}$ s$^{-1}$ (0.3--8 keV, assuming a power law model with n=2 typical for AGN), which translates to an upper limit for the host X-ray luminosity of $\sim$5$\times$10$^{40}$ erg s$^{-1}$ (a blackbody model with kT=0.1 keV results in an upper limit of 1.5$\times$10$^{40}$ erg s$^{-1}$). 

We used the {\tt uvotsource} task to construct UVOT light curves, using a 5$^{\prime\prime}$ aperture in all filters to estimate the source brightness. Background levels were estimated by using a circular region with radius of 50$^{\prime\prime}$ centered on a nearby empty region of sky. 

\subsection{Optical spectroscopy}
Optical spectroscopic observations were obtained with the New Technology Telescope (NTT) located at La Silla, Chile using the ESO Faint Object Spectrograph and Camera (EFOSC2; \citealt{Buzzoni1984}) instrument with the gr11 and gr13 grisms in combination with a 1 arcsec slit. All observations were obtained as part of the ePESSTO program. We present the observing log including observing dates, setups and exposure times in Table \ref{tab:obslog}.
 \begin{table}
 \caption{Observational setups, observing dates and exposure times of the optical long-slit EFOSC2 spectra of AT~2018fyk. A 1 arcsec slit was used for all observations. The mean MJD is given for observations taken within the same night.}
 \begin{tabular}{ccccc}
 \hline 
Grating & Obs date & MJD & Seeing & Exposure time \\\hline\hline
Gr11 & 2018--09--16 & 58\,377.112 & 1\farcs1 & 2x1800s \\
Gr11 & 2018--10--03 & 58\,394.213& 1\farcs1 & 2x1800s \\ 
Gr11 & 2018--10--18 & 58\,409.097 & 1\farcs2 & 2700s \\
Gr11 & 2018--11--01 & 58\,423.071 & 1\farcs1 & 2700s \\
Gr11 & 2018--11--15 & 58\,437.060 & 0\farcs7 & 2x2400s \\
Gr11 & 2018--12--03 & 58\,455.141 & 1\farcs1 & 2x2400s \\
Gr13 & 2018--12--16 & 58\,468.090 & 1\farcs2 & 2700s \\
Gr13 & 2018--12--17 & 58\,469.059 & 1\farcs1 & 2700s \\ 
Gr13 & 2019--01--01 & 58\,484.030 & 0\farcs8 & 2700s \\ 
Gr11 & 2019--01--09 & 58\,492.061 & 0\farcs9 & 2700s \\ 
 \hline
 \end{tabular}
   \label{tab:obslog}
 \end{table}
 \begin{figure*}
\includegraphics[width=\textwidth, keepaspectratio]{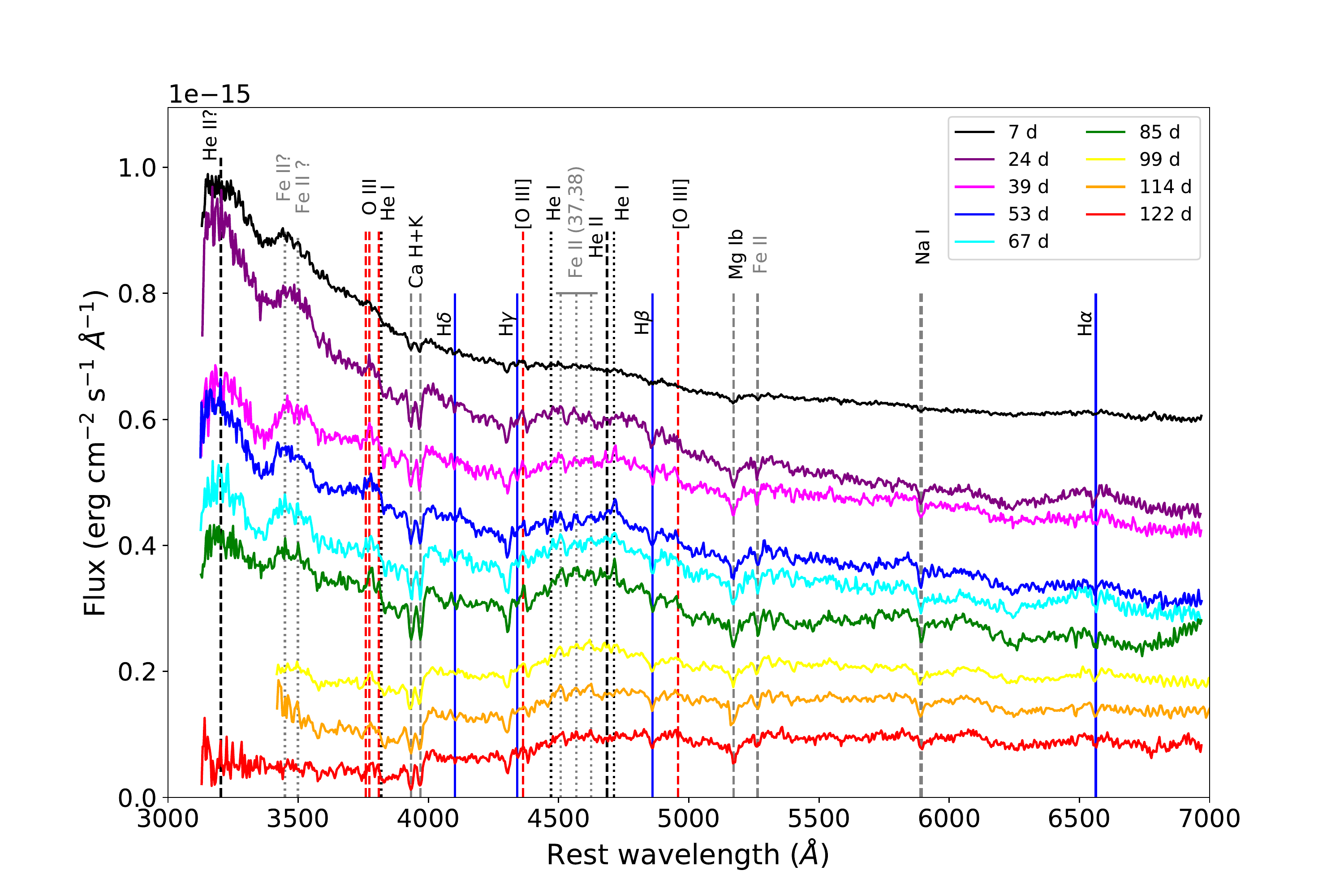}
\caption{Spectral sequence of AT~2018fyk taken with the NTT. Emission lines are marked by vertical lines: H Balmer series (solid blue), He\,\textsc{ii} (dashed black), He\,\textsc{i} (dotted black), [O\,\textsc{iii}] (solid red) and Fe\,\textsc{ii} (dotted grey). Host galaxy lines such as Ca H+K, Mg\,\textsc{i} b and Na D absorption lines are marked by dashed grey lines. The epochs are given with respect to the discovery epoch.}
\label{fig:spectra}
\end{figure*}

We reduce the spectroscopic data with \textsc{iraf}. Standard tasks such as a bias subtraction and flat field correction are performed first, after which we optimally extract the spectra \citep{Horne1986} and apply a wavelength calibration using HeAr arc lamp frames. The typical spectral resolution obtained with the gr11 and gr13 setups and a 1 arcsec slit for slit-limited observing conditions is R$\sim$ 250 and 190 at 4000 \AA, respectively (but see Table \ref{tab:obslog} for the average conditions of each observation; in seeing-limited conditions the resolution increases linearly with the average seeing). Standard star observations are used to perform the flux calibration and correct for atmospheric extinction. Given that the Galactic extinction along the line of sight is negligible, we do not try to correct for this effect. Finally, a telluric correction based on the standard star observations is applied to remove atmospheric absorption features. This is particularly useful to remove the $\lambda$\,6800 \AA\ absorption features located in the blue wing of the H$\alpha$ emission line profile. Multiple spectra taken on the same night are averaged, with weights set to the overall SNR ratio between the spectra. The spectra taken on 2018--12--16 and 2018--12--17 are also averaged due to the relatively low SNR of individual exposures. The resulting spectra are shown in Figure \ref{fig:spectra}, where the flux levels have been scaled to improve the readability of the plot. 

\subsection{Radio observations}
\begin{table}
 \caption{ATCA radio observations of AT~2018fyk. We report the time range that the array was on source, and the MJDs of the midtimes of the observations.  Flux density upper limits are obtained by stacking both frequency bands together, and are given at the $3\sigma$ level.}
 \begin{tabular}{ccccc}
 \hline 
Date & Time &  MJD & Config. & Flux density\\
& (UT) & & & ($\mu$Jy) \\
\hline
\hline
 2018-09-19 & 12:36--19:53 & 58\,380.68 & 750C & $<38$\\
 2018-10-16 & 10:04--13:28 & 58\,407.49 & 6A & $<74$\\
 2018-11-22 & 05:24--08:28 & 58\,444.29 & 6B & $<53$\\\hline
 \end{tabular}
   \label{tab:radioobs}
 \end{table}

We observed AT~2018fyk with the Australia Telescope Compact Array (ATCA) over three epochs between 2018 September 19 and 2018 November 22, under program code C3148. The observations were taken in the 750C, 6A, and 6B configurations, respectively (see Table~\ref{tab:radioobs}). While all three are east-west configurations, the former has the inner five antennas at a maximum baseline of 750\,m, with the sixth antenna located some 4.3\,km away. This isolated antenna was therefore not used when imaging the first epoch, due to the possibility of artifacts arising from the large gap in {\it uv}-coverage. In all cases we observed in the 15-mm band, using two 2048-MHz frequency chunks (each comprising 2048 1-MHz channels) centred at 16.7 and 21.2\,GHz. We used the standard calibrator PKS 1934-638 \citep{Bolton1964} as a bandpass calibrator and to set the flux density scale. To solve for the time-dependent complex gains, we used the extragalactic calibrator source QSO B2311-452 \citep[4.23$^{\circ}$ away;][]{Veron-Cetty1983} in the first epoch, and QSO B2227-445 \citep[3.29$^{\circ}$ away;][]{Savage1977} for the two subsequent epochs, as appropriate for the relevant array configurations. We used the Common Astronomy Software Application \citep[\textsc{casa} v.5.1.2;][]{McMullin2007} package for both calibration and imaging of the data, applying standard procedures for ATCA data reduction. AT~2018fyk was not detected in any of the three epochs, with upper limits as given in Table~\ref{tab:radioobs}.

\section{Analysis}
\label{sec:analysis}
\begin{figure*}
\includegraphics[width=\textwidth, keepaspectratio]{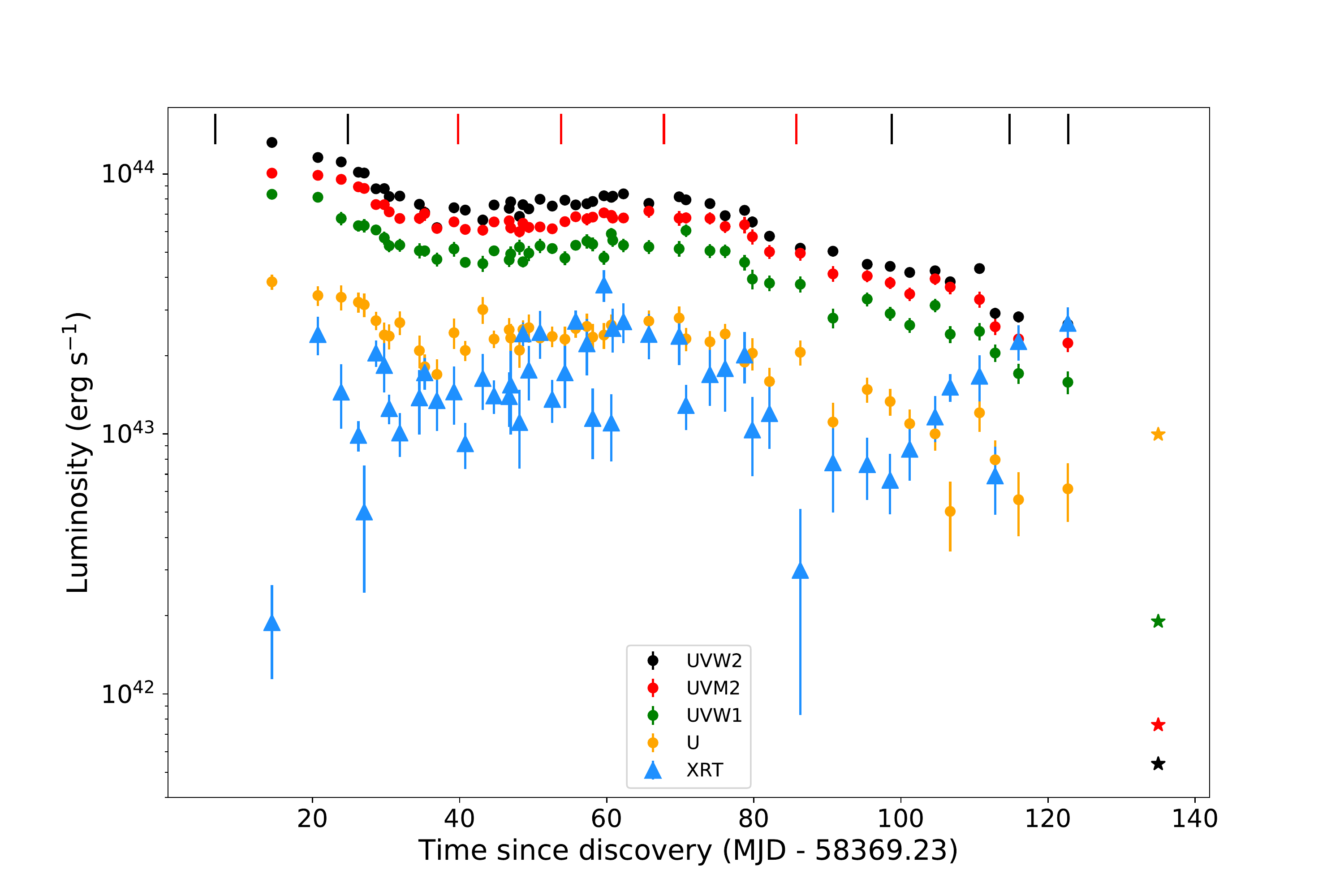}
\caption{X-ray and host-subtracted UV/optical lightcurve of AT~2018fyk as observed with {\it Swift}. The $B$- and $V$-bands are strongly contaminated by the host galaxy and are omitted for clarity. Unlike other TDE lightcurves, the UV/optical bands show a plateau phase lasting $\sim$50 days instead of a steady, monotonic decline. The dashed lines indicate epochs of spectral observations; red dashed lines indicate the epochs showing narrow emission features. Stars indicate the estimated host galaxy brightness.}
\label{fig:lc}
\end{figure*}

We present the X-ray and (host subtracted) UV/optical lightcurve obtained with {\it Swift} in Figure \ref{fig:lc}. After an initial decline, the UV/optical appears to turn over around 40 days after discovery to a near constant luminosity. This plateau lasts for nearly 50 days, before the UV/optical lightcurve breaks again to start declining, while the X-rays increase in brightness. 
While the flare is still more than 2.5 mag brighter than the host in the UV bands during the last {\it Swift} epoch, emission in the $B$- and $V$-bands was significantly contaminated by the host galaxy light even at the earliest epochs.

\subsection{SED analysis}
\begin{figure}
\includegraphics[width=0.5\textwidth, keepaspectratio]{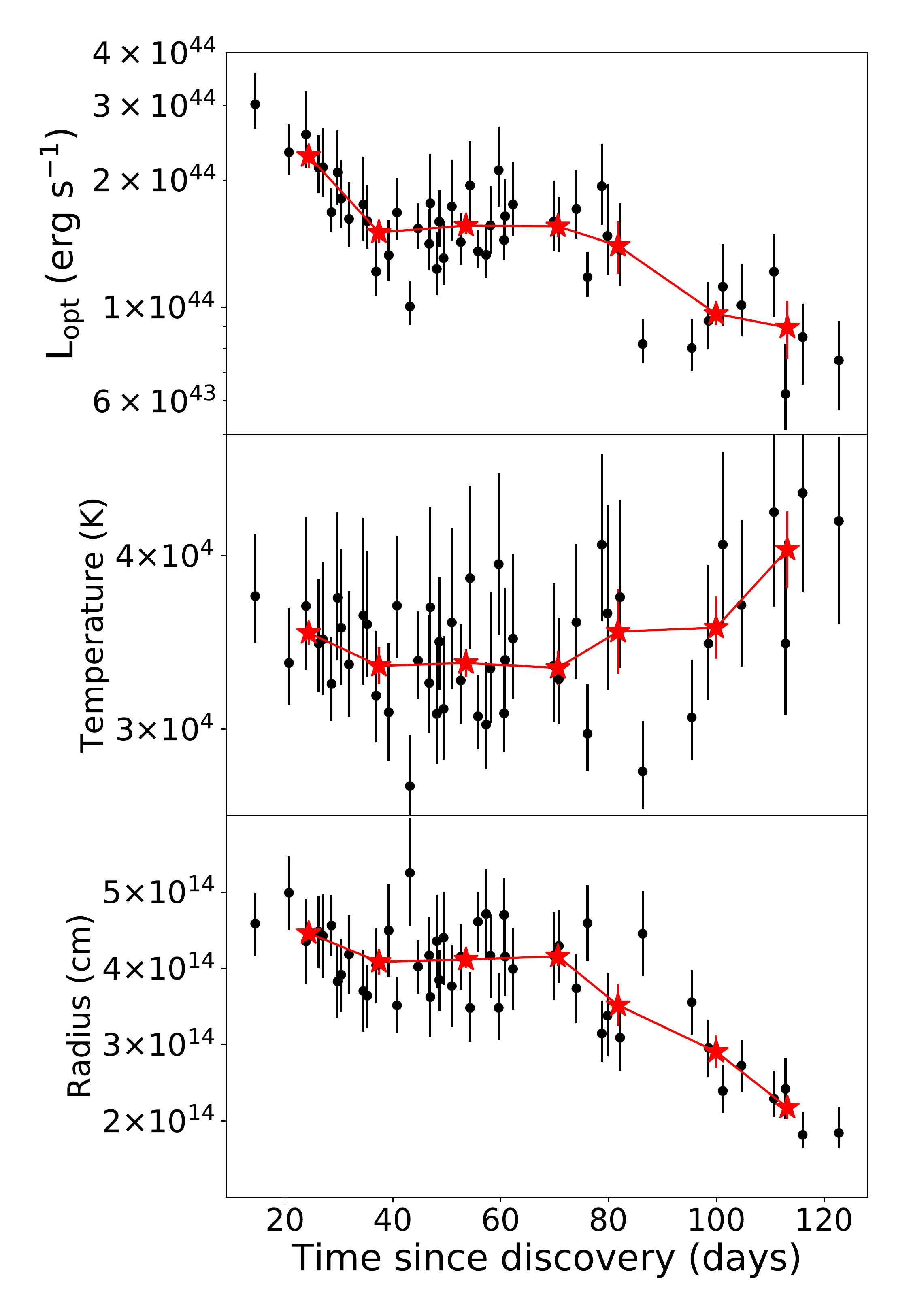}
\caption{Luminosity, temperature and blackbody radius evolution of the UV/optical component (black circles) of AT~2018fyk as derived from SED blackbody fitting and {\it Swift} XRT spectral fitting, respectively. We also show the 15 day binned lightcurves in red stars.}
\label{fig:sedanalysis}
\end{figure}

We constrain the luminosity, temperature and radius evolution of AT~2018fyk by fitting a blackbody to the {\it Swift} UVOT SED at each epoch. Due to significant contamination from the host galaxy in the reddest bands ($B$ and $V$), we do not include these data points. Including these bands does not alter the general results of our analysis, but leads to bad fits and unrealistic temperatures at some epochs. We therefore fit a blackbody model to the host subtracted SED consisting of the UV bands and the $U$-band, using a maximum likelihood approach and assuming a flat prior for the temperature between 1--5 $\times$10$^4$ K. 1 $\sigma$ uncertainties are obtained through Markov Chain Monte Carlo simulations \citep{Foreman2013}. Using the best-fit temperature at each epoch, we integrate under the blackbody curve from EUV to IR wavelengths to estimate the bolometric UV/optical luminosity. In addition, we also derive the characteristic emission radius at each epoch. We present the integrated UV/optical luminosity, temperature and radius evolution in Figure \ref{fig:sedanalysis}, both the epoch measurements and a 15 day binned evolution for clarity. We find a peak luminosity of 3.0$\pm$0.5$\times$10$^{44}$ erg s$^{-1}$, which declines by a factor of 4 over the first 120 days of the flare evolution. The temperature appears roughly constant initially, but there is evidence for an increase at later epochs similar to ASASSN--15oi \citep{Holoien201615oi} and AT2018zr \citep{Holoien2018,vanvelzen2018zr}. The radius, on the other hand, stays constant for the first $\sim$70 days at 4.2$\pm$0.4$\times$10$^{14}$ cm, after which it decreases by a factor of 2 in the span of 50 days. 
Integrating over the period with {\it Swift} coverage, we find a total UV/optical energy release of E$_{\rm rad}$ $\sim$ 1.4$\times$10$^{51}$ erg, with the uncertainties dominated by the host subtraction (the observed energy radiated at X-ray wavelengths is $\sim$10$^{50}$ erg). These values are all typical when compared to the UV/optical sample of known TDEs (e.g. \citealt{Hung2017, Wevers2017, Wevers2019, Holoien2018}).

\subsection{X-ray evolution}
AT~2018fyk belongs to a growing sample of UV/optical detected TDE candidates observed to be X-ray bright at early times, together with ASASSN--14li \citep{Holoien201614li}, ASASSN--15oi \citep{Holoien201615oi} and PS18kh/AT2018zr \citep{Holoien2018, vanvelzen2018zr}. In addition the source XMMSL1 J0740 \citep{Saxton2016} was also UV/optical and X-ray bright, although it was detected in X-rays first. 

The {\it Swift} XRT lightcurve shows variability of a factor 2--5 on a timescale of days, whereas the UV/optical lightcurve appears more smooth. During the first epoch the source shows a L$_{\rm opt}$/L$_{\rm x}$ ratio $\sim$150, similar to that of ASASSN--15oi \citep{Gezari2017} and AT2018zr \citep{Holoien2018,vanvelzen2018zr}. The X-ray emission then brightens by a factor of $\sim$10 in 6 days, and remains roughly constant for 25 days. The X-ray emission then displays a plateau similar to the UV/optical evolution, leading to a near constant L$_{\rm opt}$/L$_{\rm x}$ ratio for $\sim$70 days. Between 80 and 100 days after discovery the lightcurves decline in tandem, after which the X-rays brighten once more while the UV/optical emission keeps declining. 

\begin{figure}
\includegraphics[width=0.5\textwidth, keepaspectratio]{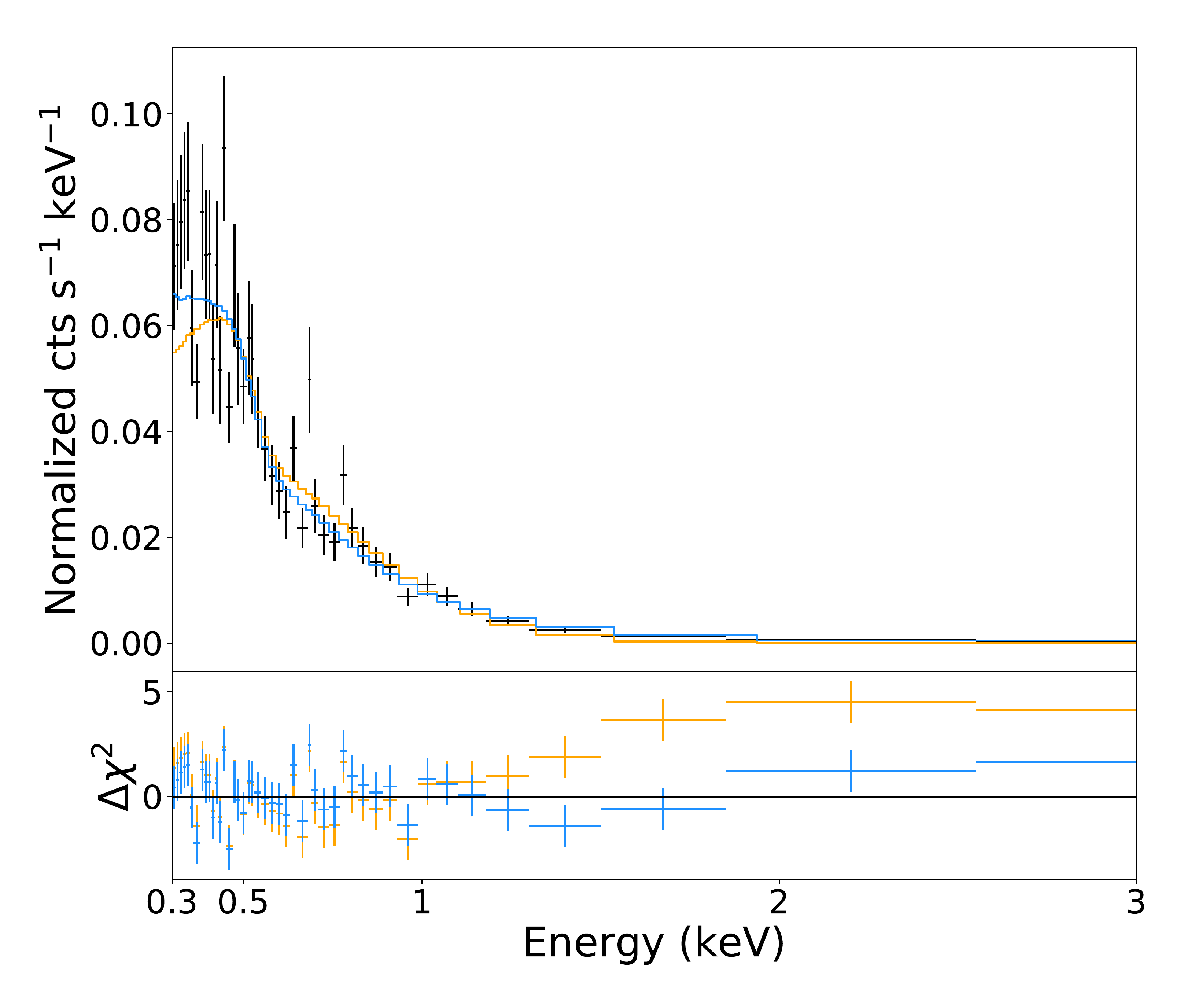}
\caption{Stacked, rebinned 50.2 ks X-ray spectrum obtained with {\it Swift}. We overplot the best-fit absorbed multi-temperature blackbody model (\textsc{diskbb} in xspec, orange) with T$_{\rm in}$\,=\,162$\pm$4 eV and negligible n$_{\rm H}$. The residuals show that the flux is systematically underestimated at energies $>$1.5 keV, and an absorbed power-law + blackbody model (blue) describes this higher energy emission better.}
\label{fig:xrayspec}
\end{figure}

We first rebin the stacked spectrum (total exposure time of 50.2 ks, Figure \ref{fig:xrayspec}) to obtain at least 25 counts per spectral bin, appropriate for the use of $\chi^2$ statistics in \textsc{xspec}. Fitting this spectrum with a blackbody model (tbabs$\times$zashift$\times$bbodyrad in \textsc{xspec}), we find a best-fit temperature ($\chi^2$=3.62 for 43 degrees of freedom [dof]) of kT=121$\pm$2 eV, negligible n$_{\rm H}$ and a normalisation factor norm=675$\pm$63. This normalisation corresponds to a X-ray photospheric radius of R$_{\rm X}$=6.5$\pm$0.3$\times$10$^{10}$ cm. This in turn corresponds to the innermost stable circular orbit of an accretion disk around a non-spinning SMBH of $\sim$10$^5$ M$_{\odot}$. This is a factor of $\sim$100 lower than inferred from the bulge mass, and suggests that some obscuration (either from tidal debris or in the host galaxy) occurs. Given the high reduced $\chi^2$ of the fit, we also try an absorbed multi-temperature blackbody model (tbabs$\times$zashift$\times$diskbb) and find T$_{\rm in}$=162$\pm$4 eV ($\chi^2$=2.77 for 43 dof). From Figure \ref{fig:xrayspec} (the orange line and markers) it is clear that an additional emission component at energies $>$1.5 keV is required. Using an absorbed power-law + blackbody model (phabs$\times$zashift$\times$(powerlaw+bbodyrad)) increases the goodness of fit significantly ($\chi^2$=1.39 for 41 dof); this model is shown in blue in Figure \ref{fig:xrayspec}. The power-law component contributes $\sim$\,30 \% to the unabsorbed X-ray flux. Further analysis is required to investigate the detailed spectral evolution and the potential presence of a harder emission component similar to XMMSL1 J0740 \citep{Saxton2016}, but we defer a more detailed temporal and spectral analysis of the X-ray data to a companion paper. 


\subsection{Optical spectroscopy}
The earliest epochs of spectroscopic observations are dominated by a hot, featureless continuum with several broad emission lines superposed. 
We identify broad H$\alpha$, \hetwo $\lambda$4686 and potentially \hetwo 3203 \AA\ emission lines in the spectrum. In addition, we identify a broad emission line (or lines) in the region 3400--3600\AA.
This latter feature can be tentatively identified as O\,\textsc{iii} $\lambda$3444 or potentially broad Fe\,\textsc{ii} ($\lambda\lambda$3449,3499) lines, although these identifications are uncertain. 

AT~2018fyk became unobservable due to Sun constraints before the broad emission lines completely disappeared, hence we cannot perform the host galaxy subtraction in the traditional way. Instead, to identify the nature of the lines and measure their line widths and velocity offsets, we first fit cubic splines to the continuum in \textsc{molly}\footnote{\textsc{molly} is an open source spectral analysis software tool.}, masking all prominent emission features, host and remaining telluric absorption lines. We then subtract the continuum level to reveal the TDE emission line spectrum. Although some host contamination remains, in particular narrow absorption lines (such as the H\,$\beta$ absorption trough in the red wing of He\,\textsc{ii}), to first order this removes the featureless blackbody and host galaxy continuum contributions.

\begin{figure*}
\includegraphics[width=\textwidth, keepaspectratio]{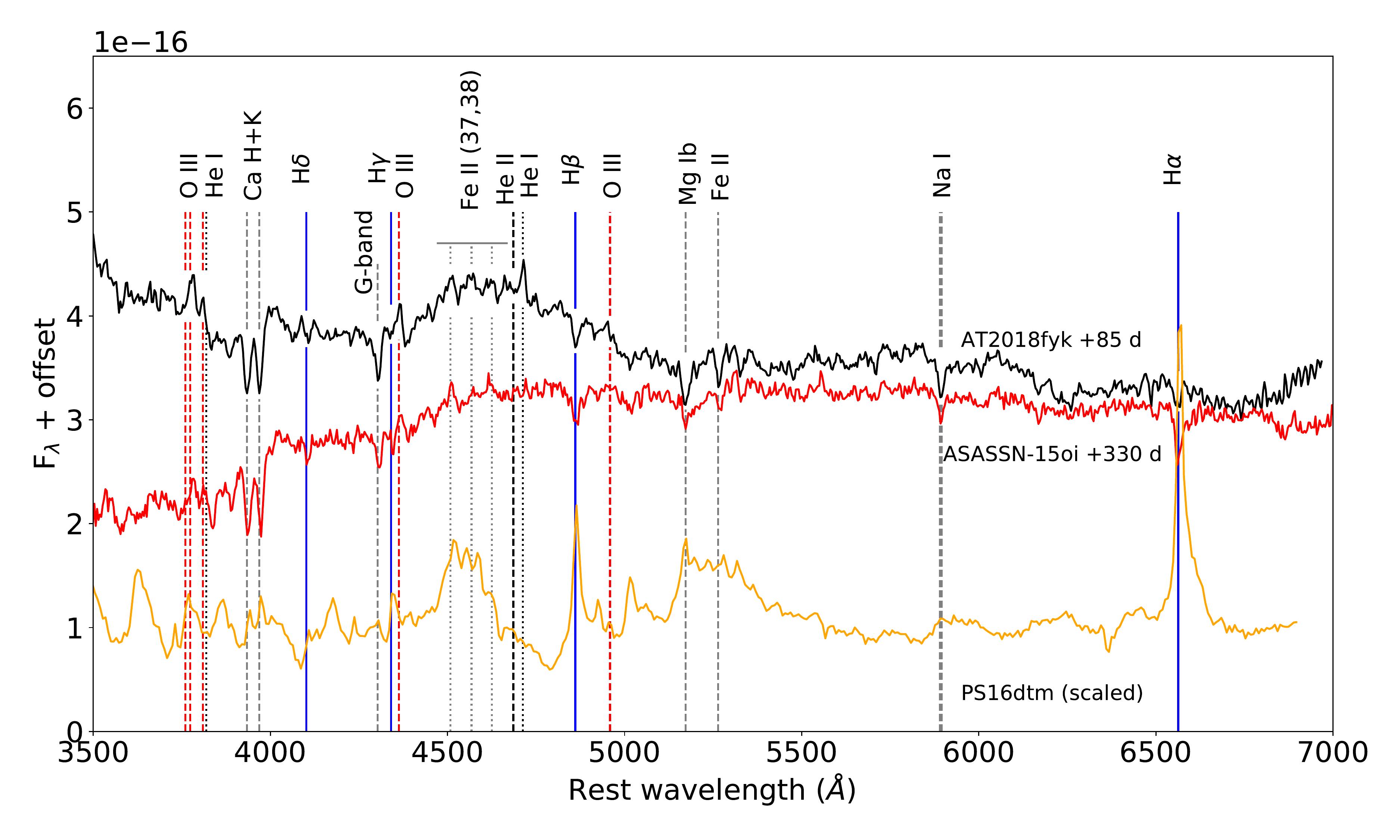}
 \caption{Comparison of the emission line profiles in the \hetwo 4686 region with other events. The narrow Fe\,\textsc{ii} lines are indicated by vertical lines dotted lines. We show the ASASSN--15oi spectrum in which we identify similarly narrow Fe\,\textsc{ii} lines in red. We also show the spectrum of PS16dtm, which showed very strong Fe\,\textsc{ii} emission.}
\label{fig:spectra1}
\end{figure*}

\begin{figure*}
\includegraphics[width=0.49\textwidth, keepaspectratio]{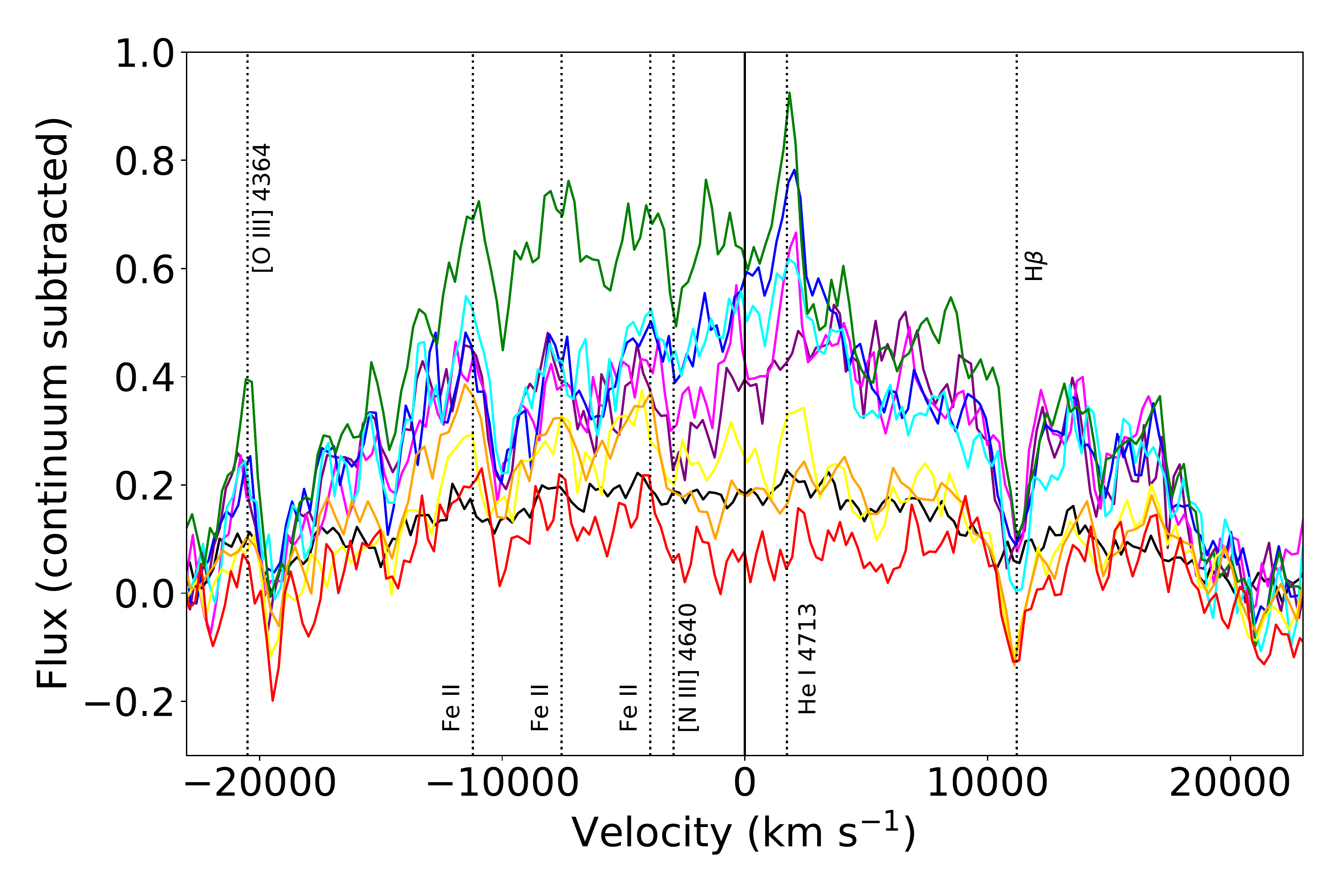}
\includegraphics[width=0.49\textwidth, keepaspectratio]{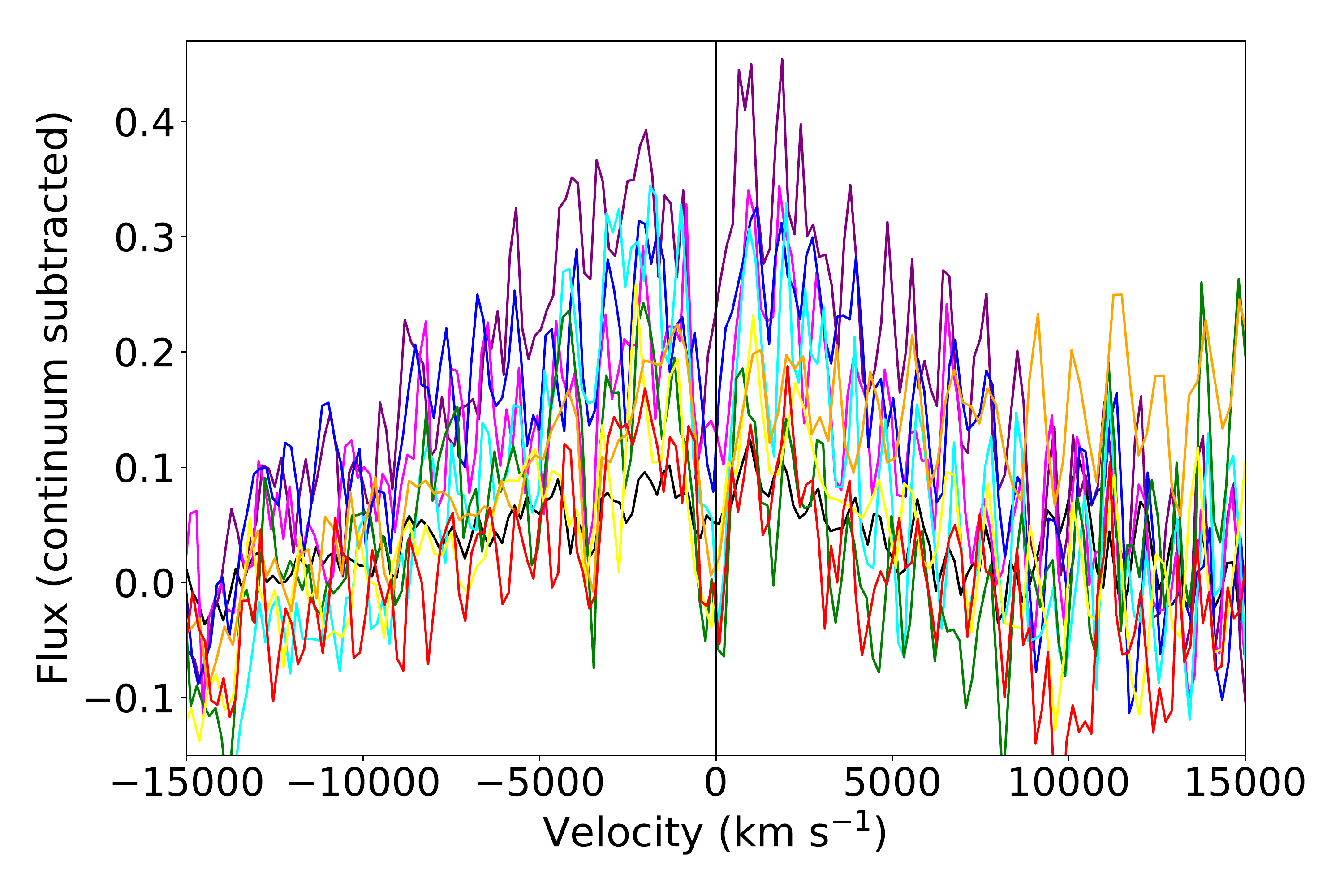}
 \caption{The markedly distinct line profile evolution of the \hetwo 4686 complex and H$\alpha$ (left and right panels, respectively). Relevant emission and absorption lines are marked by vertical lines.}
\label{fig:velspace}
\end{figure*}

Arguably the most interesting features in the optical spectra of AT~2018fyk are the narrow emission lines that appear after the lightcurve shows a plateau in luminosity. We show the spectrum with the most prominent narrow emission features in Figure \ref{fig:spectra1}, including the most likely line identifications. We identify several high ionisation O\,\textsc{iii} lines, and in addition we identify several low ionisation Fe\,\textsc{ii} emission lines (ionisation potential $\sim$ 8 eV), particularly of the multiplets 37 and 38 with prominent features at $\lambda\lambda$4512,4568,4625. We also identify low excitation He\,\textsc{i} narrow emission lines. Moreover, the increased pseudo-continuum level in Figure \ref{fig:velspace} (the green spectrum) may suggest that the emergence of these narrow Fe\,\textsc{ii} lines is accompanied by a broad component as seen in AGN, although this could also be the forest of narrow Fe\,\textsc{ii} lines that is present in the wavelength range 4300--4700 \AA. This shows that the spectral diversity of TDEs is even larger than previously identified, with a class of Fe-rich events in addition to the H-, He- and N-rich TDEs \citep{Arcavi2014,Hung2017, Blagorodnova2017,Leloudas2019}.

The fact that the narrow emission lines are observed only when the lightcurve shows a plateau phase strongly suggests that they are powered by the same emission mechanism.
We also note that we only see narrow emission lines in the blue part of the spectrum. Several transitions of both He\,\textsc{i} and O\,\textsc{iii} exist at longer wavelengths, and these transitions typically have stronger line strengths (for example in AGN) than the lines we observe in AT~2018fyk. 

\section{Discussion}
\label{sec:discussion}
\subsection{TDE classification}
We classified AT~2018fyk as a TDE candidate based on several pieces of evidence. 

First, the location is consistent to within $\sim$100 pc with the nucleus of a galaxy. No signs of activity or star formation are evident from the galaxy colours and no narrow galaxy emission lines are present in the spectra, arguing against a supernova interpretation. Archival X-ray upper limits show that the X-ray emission brightened by a factor of $\gtrsim$1000, making an AGN flare an unlikely interpretation.

Second, the temperature, colour and blackbody radius evolution of the UV/optical emission are typical of TDEs and unlike any other known SN types \citep{Hung2017,Holoien2018}. The optical spectral evolution is also unlike any SN spectra.

Third, the X-ray emission is an order of magnitude brighter than the brightest X-ray supernovae observed (e.g. \citealt{Dwarkadas2012}), including superluminous supernovae \citep{Margutti2018}. Moreover, the X-ray spectra of supernovae are not expected to be well described by soft thermal blackbody emission.

The observed properties are broadly consistent with observed TDEs: hot (T$\sim$3.5$\times$10$^{4}$ K) UV/optical blackbody emission that does not cool over 100 days, a near-constant UV/optical colour evolution, a thermal blackbody X-ray component with a temperature of $\sim$\,100 eV, broad ($\sim$\,15000 km s$^{-1}$) H and He optical emission lines can all be naturally explained in the TDE scenario \citep{Arcavi2014, Hung2017, Blagorodnova2017, Wevers2017, Holoien2018}. 
In the remainder of this Section, we discuss several peculiar features (compared to observations of other TDEs) and how they can be explained in the TDE interpretation.

\subsection{Lightcurve comparison and secondary maxima}
\begin{figure}
\includegraphics[width=0.5\textwidth, keepaspectratio]{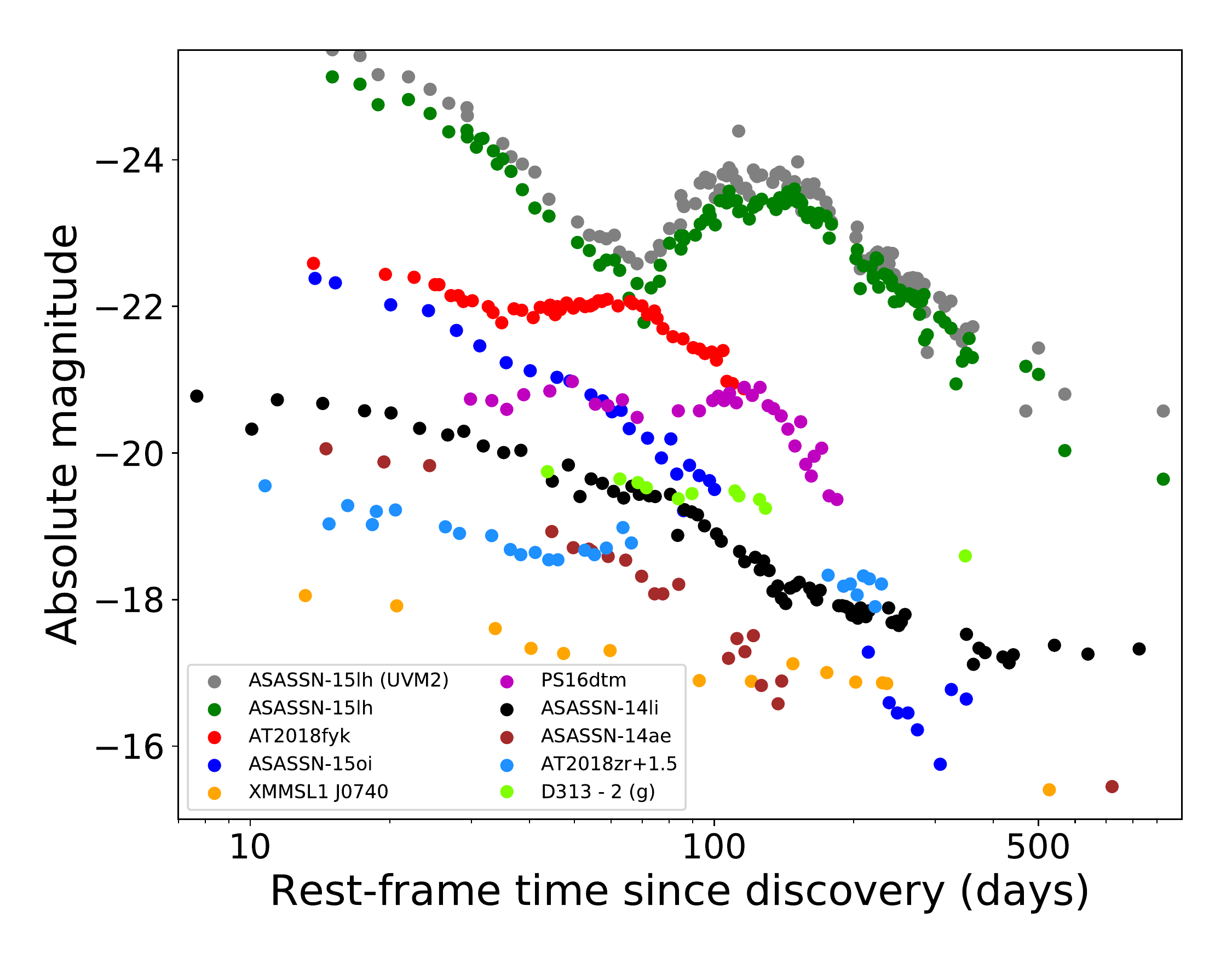}
 \caption{Comparison of the AT~2018fyk UVW2 lightcurve with other TDEs and TDE candidates near the UV/optical peak. A secondary maximum similar to ASASSN--15lh is observed. In addition, several other sources including AT~2018zr, XMMSL1 J0740 as well as ASASSN--14li show a clear secondary maximum in their lightcurve.}
\label{fig:lccomp}
\end{figure}
To put the lightcurve shape into context, in Figure \ref{fig:lccomp} we compare the UVW2 lightcurve of AT~2018fyk with other TDE candidates. The $V$-band (observed) peak absolute magnitude is V = --20.7. While the decline is monotonic for the first $\sim$40 days, similar to the TDEs ASASSN--14li and ASASSN--15oi, the lightcurve plateaus before declining at a rate similar to ASASSN--14li. This is reminiscent of the behaviour seen in the TDE candidate ASASSN--15lh, which shows a similar (albeit much more pronounced and much longer) secondary maximum in its lightcurve. Given the much higher redshift of the latter source, we also show its UVM2 lightcurve, which probes similar rest wavelengths to the UVW2 filter for the other events. The lightcurve of PS16dtm, which has been claimed to be a TDE in a NLS1 galaxy \citep{Blanchard2017}, shows a plateau phase but not the characteristic decline from peak leading up to it, as seen in AT~2018fyk and ASASSN--15lh.
The UVW2 lightcurve of XMMSL1 J0740 also shows a similar, though less pronounced, rebrightening phase at $\sim$150 days \citep{Saxton2016}. 
For the TDE candidate ASASSN--15lh (but see e.g. \citealt{Dong2016, Bersten2016} for an extreme supernova interpretation), \citet{Leloudas2016} propose that the rebrightening can be explained by taking into account the SMBH mass, which is by far the most massive of the TDE sample ($>$ 10$^8$ M$_{\odot}$). As a consequence, all orbital pericenters become relativistic, even for shallow (low $\beta$=R$_{\rm T}$/R$_{\rm p}$) stellar encounters. 
Similar to ASASSN--15lh, we propose that a relativistic pericenter, which leads to two peaks in the lightcurve \citep{Ulmer1999}, can explain the observations; the first maximum due to shock energy released during stream self-intersections, and the second after disk formation, powered by accretion onto the SMBH\footnote{Recent work \citep{Bonnerot2019}, which appeared while this manuscript was under review, suggests that the radiative efficiency of the stream-stream shock could be too low to explain the peak luminosity of observed TDEs (including AT~2018fyk). In the context of their model, the first peak of the light curve could be powered by a "secondary shock" at the trapping radius, while the plateau is caused by subsequent accretion.}. The relatively short timescale between the first and second maxima in the lightcurve favours a star from the lower end of the stellar mass distribution, which decreases the semi-major axis and orbital time of the most bound stellar debris (e.g. \citealt{Dai2015}).

We do not have an accurate black hole mass measurement for the host of AT~2018fyk, but a rough estimate based on the stellar population synthesis suggests that M$_{\rm BH}$ $\sim$ 2\,$\times$\,10$^7$ M$_{\odot}$. Using a simple theoretical prediction of the peak fall-back rate \citep{Stone2013}, this will lead to sub-Eddington fall-back rates (and hence luminosities, as observed), similar to other TDEs with high black hole masses (e.g. TDE1 and D3-13, \citealt{Wevers2017}). Given this relatively high black hole mass, for a sub-Eddington peak fall-back rate and in the presence of strong shocks during stream self-intersection due to the relativistic pericenter, it is expected that disk formation is more efficient than for non-relativistic pericenters. This holds true for all TDEs around black holes $\gtrsim$10$^7$ M$_{\odot}$, so we inspect the lightcurves of TDE1 and D3-13 for similar signatures. While the lightcurve for TDE1 is very sparsely sampled, relatively good coverage is available for D3-13. We find evidence for a rebrightening in the g-band lightcurve $\sim$100 days after observed peak, as well as a marked flattening in the r- and i-band lightcurves. The effects are likely to be strongest at UV wavelengths, which are not covered for D3-13. Nevertheless this suggests that a double-peaked lightcurve could be a quasi-universal signature of TDEs around massive ($>$10$^7$ M$_{\odot}$) black holes, and observations of future TDEs with such black hole masses can confirm this. This interpretation is also consistent with the observed SMBH mass dependence of the late-time UV excess \citep{vanvelzen2019}, where TDEs around higher mass SMBHs have no late-time excess because the early-time emission already includes a large disk contribution due to more efficient circularization.

Our M$_{\rm BH}$ estimate was obtained using scaling relations different from the M\,--\,$\sigma$ relation, and the estimate could potentially be revised downward by up to an order of magnitude (similar to other TDE hosts with M$_{\rm BH}$ estimates from both the M\,--\,L and M\,--\,$\sigma$ relations). In that case, the peak fall-back rate and luminosity might be super-Eddington and Eddington limited, respectively, and the scenario outlined above becomes unlikely (unless the encounter had a high impact parameter to make the pericenter relativistic). Instead, a variable super-Eddington disk wind (which quenches as the fall-back rate decreases) could explain the reprocessing of X-rays into UV/optical emission. When the accretion rate drops further, the disk transitions into a thin disk state, increasing the viscous timescale and flattening the lightcurve. We note that a super-Eddington luminosity is not necessarily required for this scenario, as disk transitions can occur even at a few $\times$0.1 L$_{\rm Edd}$ (e.g. \citealt{Abramowicz1988}), which is plausible for AT~2018fyk. A velocity dispersion measurement for the host SMBH is required to more accurately measure the black hole mass and differentiate between these scenarios.

Below, we will argue that the second peak in the lightcurve is powered by efficient reprocessing of energetic photons from the central source into UV/optical emission.
As a final note, identifying a TDE candidate with more typical TDE host galaxy parameters \citep{Wevers2019} but observational characteristics similar to ASASSN--15lh argues in favour of the TDE interpretation of that event (as opposed to a unique SN interpretation). In this interpretation the UV/optical emission and the emergence of X-ray emission after an initial non-detection are explained by the rapid formation of an accretion disk. These similarities and the link between the UV/optical and X-ray emission strengthen the classification of ASASSN--15lh as a TDE \citep{Leloudas2016, Margutti2017}.

\subsection{Detection of low ionisation, narrow emission lines}
\begin{figure*}
\includegraphics[width=\textwidth, keepaspectratio]{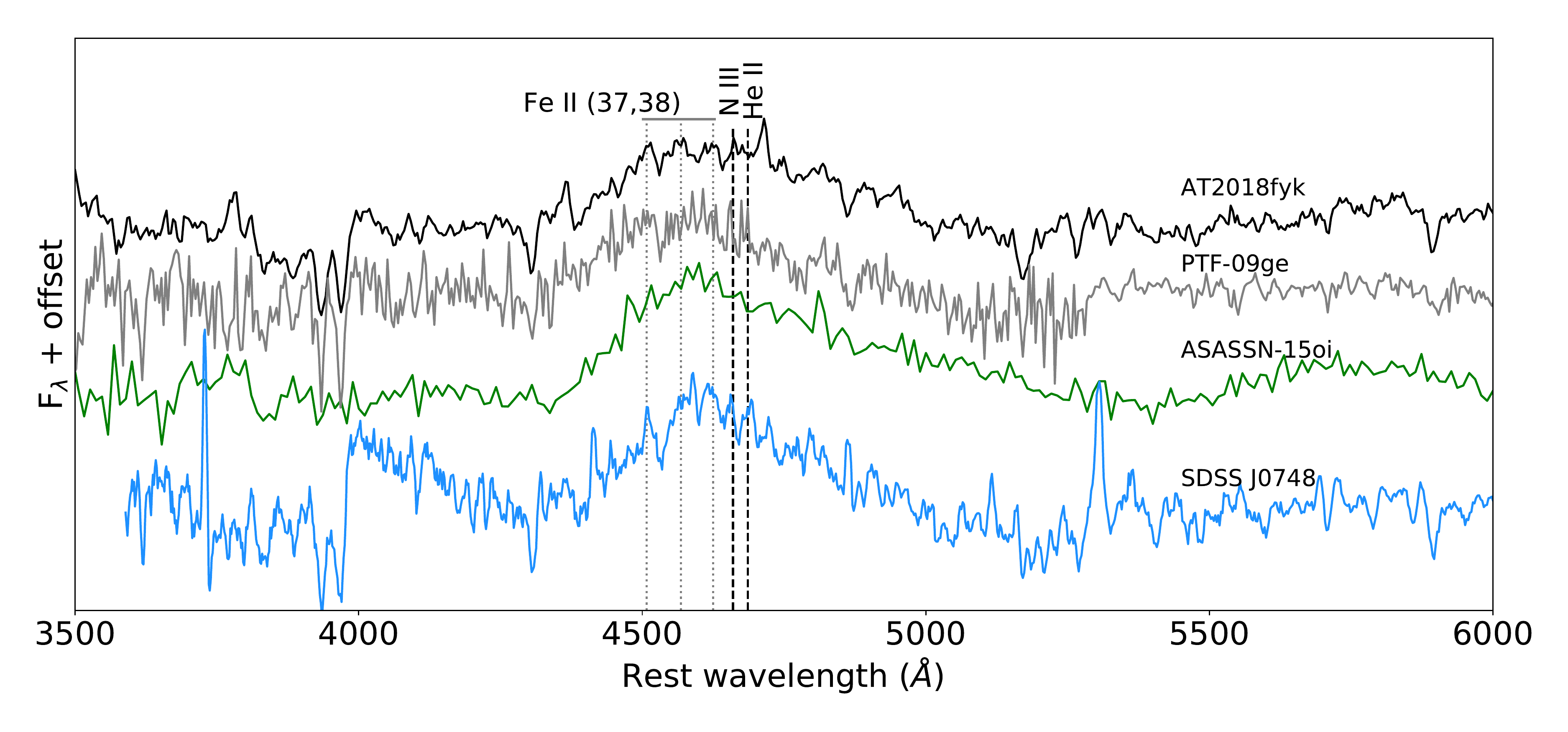}
 \caption{Spectral comparison of AT~2018fyk with ASASSN--15oi, PTF--09ge and SDSS J0748. All events display a distinct asymmetric line profile in the region around He\,\textsc{ii} $\lambda$ 4686, which we propose can be explained by multiple Fe\,\textsc{ii} emission lines.}
\label{fig:comparison}
\end{figure*}

The broad emission feature near 4686\,\AA, if only associated with  \hetwo 4686 emission, is non-Gaussian in several of the spectra. Comparing its FWHM $\sim$28000 km s$^{-1}$ with that of the other lines, which range between 10--15 $\times 10^3$ km s$^{-1}$, it is hard to explain why this line is almost twice as broad if it originates in roughly the same physical region. Moreover, the line develops a distinct asymmetric blue shoulder during its evolution (Figures \ref{fig:velspace} and \ref{fig:comparison}). 
This suggests, as has been noticed in other TDEs (e.g. \citealt{Arcavi2014, Holoien201615oi, Leloudas2019}) that instead this line might be a superposition of several emission features. \citet{Holoien201615oi} suggested that part of this line might be explained by \heone 4472 in ASASSN--15oi; for AT~2018fyk the line would be redshifted by $\sim$2500 km s$^{-1}$, which is not observed for H$\alpha$ and He\,\textsc{ii}.  \citet{Leloudas2019} explain the asymmetry in some TDEs as a consequence of Bowen fluorescence lines, but we do not observe the characteristic N\,\textsc{iii} $\lambda\lambda$4097,4103 feature that is expected in this case. This suggests that in AT~2018fyk and potentially other TDEs such as ASASSN--15oi (see fig. 4 in \citealt{Leloudas2019}), Bowen fluorescence lines do not provide a satisfactory explanation. Another alternative, suggested by \citet{Roth2018}, is outflowing gas that is optically thick to electron scattering, which can produce blue-shifted emission peaks and asymmetric red wings in the line profiles.

The emergence of the narrow spectral lines in AT~2018fyk (Figure \ref{fig:spectra1}) allows us to identify the emission in this blue shoulder as Fe\,\textsc{ii} multiplet 37,38 emission lines. These are the strongest optical Fe\,\textsc{ii} multiplet lines, although depending on the excitation mechanism one might also expect emission in the NIR around 1$\mu$m \citep{Marinello2016}, which is unfortunately not covered by our spectra. Given the similarity of the line profiles, we propose that the origin of the blue bump near \hetwo 4686 in the other two events shown in Figure \ref{fig:comparison}, ASASSN--15oi and PTF--09ge, is likewise Fe\,\textsc{ii} emission, making these events part of an {\it Fe-rich} class of TDEs. We have also included the coronal line emitter and TDE candidate SDSS J0748 \citep{Yang2013} for comparison because the line shape is remarkably similar.

These low ionisation lines have been detected in AGNs (e.g. \citealt{Lawrence1988,Graham1996}), with EWs that can exceed those of \hetwo 4686. Although the excitation mechanism(s) in AGN is somewhat ambiguous, photo ionisation \citep{Kwan1981}, Ly$\alpha$ resonance pumping \citep{Sigut1998} and collisional excitation (depending on the particle density) have all been proposed to contribute to some extent to produce these transitions \citep{Baldwin2004}. Their strength is closely associated with the Eddington fraction in AGN \citep{Boroson1992, Kovacevic2010}. While the narrow Fe\,\textsc{ii} lines are thought to originate from a well defined region in between the broad line region (BLR) and narrow line region (NLR), the emission region of the broad component is not currently well constrained \citep{Dong2011}. One possibility is that it originates from the surface of the AGN accretion disk \citep{Zhang2006}; further evidence for an origin in the accretion disk comes from cataclysmic variables (e.g. \citealt{Roelofs2006}). Interestingly, \citet{Dong2010} showed that while optical Fe\,\textsc{ii} emission is prevalent in type 1 AGN, it is not observed in type 2 AGN. This suggests that the emission region is located within the obscuring torus.

More generally, the emission region is likely a partially ionised region, where the ionising photons come from a central X-ray source \citep{Netzer1983}.  Incidentally, some of the strongest optical Fe\,\textsc{ii} lines are observed in narrow-line Seyfert 1 (NLS1) galaxies (e.g. \citealt{Osterbrock1985}), which are typically characterised by a significant soft X-ray excess below 1.5-2 keV, rapid X-ray flux and spectral variability (see e.g. the review by \citealt{Gallo2018}) and potentially accreting at high fractions of their Eddington rate \citep{Rakshit2017}. Another interesting resemblance is their preferred black hole mass range, which is $<$10$^8$ M$_{\rm BH}$ for both TDEs and NLS1s \citep{Peterson2011, Berton2015, Chen2018}. These properties are all remarkably similar to those expected/observed for TDEs. 

In particular, the TDE candidate PS16dtm was suggested to be a TDE in an active galaxy \citep{Blanchard2017}; the spectrum resembles that of NLS1 galaxies, showing several optical Fe\,\textsc{ii} lines (Figure \ref{fig:spectra1}). PS1-10adi, another TDE candidate in an AGN, was also observed to produce transient Fe\,\textsc{ii} optical emission at late times \citep{Kankare2017}; similar features were also observed in the TDE candidates and extreme coronal line emitters SDSS J0748 and SDSS J0952 \citep{Wang2011, Yang2013}. These events all occurred around active black holes, so establishing their TDE nature is more ambiguous. The resemblance of AT~2018fyk to some of these events shows that stellar disruptions can create (temporary) circumstances very similar to those in NLS1 AGN even around dormant SMBHs, and that instead of several distinct classes there may be a continuum of nuclear transient events intermediate to `clean' TDEs and `clean' AGN flares.

Unlike the high ionisation narrow lines such as O\,\textsc{iii} (which are thought to form within the ionisation cone of the central X-ray source in AGN), Fe\,\textsc{ii} emission requires an obscuring medium with significant particle density and optical depth as well as heating input into the gas. The presence of these Fe\,\textsc{ii} lines in the spectra of AT~2018fyk indicates that at least part of the gas is optically thick, while the X-ray spectrum shows that a bright, soft X-ray source is present, making the conditions in this TDE similar to that in NLS1 nuclei.

We inspect publicly available spectra of other TDEs, and find that the presence of narrow Fe\,\textsc{ii} lines is not unique to AT~2018fyk. We identify similar emission lines consistent with the same Fe\,\textsc{ii} multiplet 37,38 lines in optical spectra of ASASSN--15oi at late phases ($\sim$330 days after discovery; Figure \ref{fig:spectra1}). 
Upon further investigation, the L$_{\rm opt}$/L$_{\rm x}$ ratio of both sources is nearly constant while the narrow lines are present (Figure \ref{fig:loptlx}; see also Section \ref{sec:loptlx}). This suggests that the L$_{\rm opt}$/L$_{\rm x}$ ratio evolution in ASASSN--15oi at late times may be similarly regulated by reprocessing of soft X-ray radiation in optically thick gas, analogous to the situation in AT~2018fyk and AGNs. 

The formation of an accretion disk that radiates in soft X-rays, which subsequently partially ionise high density, optically thick gas surrounding the SMBH delivered by the disruption can explain the emergence of the Fe\,\textsc{ii} emission lines. At the same time, the reprocessing of X-ray, Ly$\alpha$ and/or EUV photons can power the plateau phase in the lightcurve, explaining both peculiar features in the TDE scenario. \citet{vanvelzen2019} showed that the late-time plateau phase can be explained by UV disk emission, and this can also contribute to the plateau phase seen in AT~2018fyk.

While high temporal coverage in X-ray and UV/optical wavelengths is available for only a few candidates, the UV/optical lightcurve shape of AT~2018fyk is unique among UV/optical bright TDEs. If we are indeed witnessing the assembly of an accretion disk and reprocessing of disk X-ray radiation, this implies that it does not occur with a similar efficiency in most TDEs. The first 40 days of the lightcurve, however, show typical behaviour as observed in nearly all UV/optical TDEs (Figure \ref{fig:lccomp}). The plateau represents an additional emission component superposed on the contribution responsible for the initial decline from peak. \citet{vanvelzen2019} showed that such a secondary maximum is observed in nearly all TDEs, but several years after disruption rather than several months as observed in AT~2018fyk and ASASSN--15lh. 

\subsection{Broad iron emission lines?}
In terms of velocities, the \hetwo 4686 and H$\alpha$ lines follow a similar trend, being consistent with their respective rest wavelengths in early epochs but becoming more blueshifted up to about 2000 km s$^{-1}$, with a blueshift of $\sim$1000 km s$^{-1}$ in the latest spectrum. Although the \hetwo 3202 line can tentatively be identified in the spectra, it is on the edge of the spectrum and a sudden decrease in instrumental through-put may instead be responsible for this feature. More interestingly, the (broad) line that we tentatively identify as O\,\textsc{iii} at 3444 \AA\ or \heone at 3446 \AA\ seems to be systematically redshifted by 2000--3000 km s$^{-1}$. Fitting a single Gaussian profile to this line, we find central wavelengths ranging between 3375 and 3500 \AA\ during the evolution. However, the line has a rather boxy profile instead of being well described by a Gaussian. In this wavelength range, two narrow emission features with rest wavelength of 3449 and 3499 \AA\ are visible during the nebular phase (Figure \ref{fig:spectra}). While the former is consistent with either O\,\textsc{iii} 3444 \AA\ or \heone at 3447 \AA, the identification of the latter is 3499 \AA\ line is less secure. As an alternative, the NIST Atomic Spectra Database shows several strong Fe\,\textsc{ii} transitions corresponding to wavelengths close to 3449 and 3499 \AA. If these line identifications as Fe\,\textsc{ii} are correct, this provides unambiguous evidence for broad Fe\,\textsc{ii} emission lines in the early spectroscopic observations (Figure \ref{fig:spectra}).

We also tentatively identify the emergence of a broad emission feature around He\,\textsc{i} 5876\AA\ that is present in several epochs. Without a solid host galaxy subtraction, however, this feature must be interpreted with caution as it is unclear what constitutes the continuum level, given the many broad features and bumps present in the spectra. In addition, there is a deep absorption feature that distorts the line shape. We tentatively identify this feature as He\,\textsc{i} 5876\AA\ but a proper host galaxy subtraction is needed to study the line evolution in more detail.
 
 \subsection{Optical to X-ray ratio evolution}
 \label{sec:loptlx}
  \begin{figure}
\includegraphics[width=0.5\textwidth, keepaspectratio]{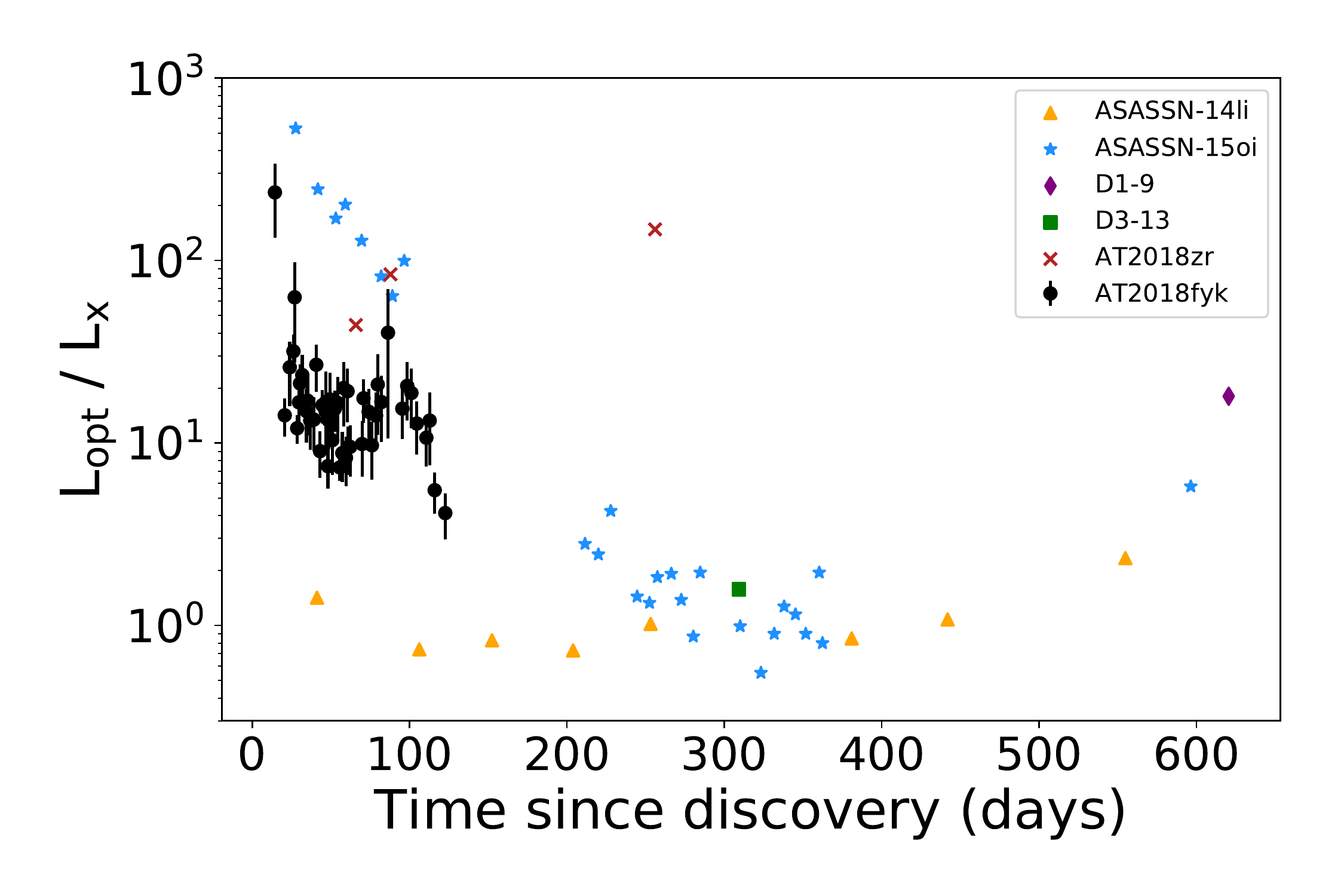}
\caption{UV/optical to X-ray luminosity ratio for AT~2018fyk, as well as several other X-ray bright TDEs. Data taken from \citet{Gezari2017} and \citet{vanvelzen2018zr}.}
\label{fig:loptlx}
\end{figure}
 AT~2018fyk is only the third TDE candidate with contemporaneous bright UV/optical and X-ray emission that has been observed by {\it Swift} with high cadence at both wavelength regimes.
We show the ratio of integrated UV/optical luminosity to X-ray luminosity in Figure \ref{fig:loptlx}, where we also overplot these ratios for ASASSN--14li, ASASSN--15oi and AT2018zr \citep{Gezari2017, vanvelzen2018zr}. The L$_{\rm opt}$/L$_{\rm x}$ ratio of ASASSN--14li is $\sim$1 for 400 days, with some hint of an increase at later times. On the other hand, the evolution of the L$_{\rm opt}$/L$_{\rm x}$ ratio of ASASSN--15oi is markedly different, and has been interpreted as the delayed formation of an accretion disk \citep{Gezari2017}.
 The evolution of AT~2018fyk appears to broadly follow that of ASASSN--15oi, as it decreases over time. However, rather than a monotonic decrease sudden changes are apparent at early times and during the 2 most recent {\it Swift} observations. 
 The L$_{\rm opt}$/L$_{\rm x}$ ratio appears to plateau for $\sim$80 days similar to the UV/optical lightcurves, after which it decreases as the X-ray luminosity brightens and the X-ray spectrum becomes harder. 

During this plateau phase, both the X-ray and UV/optical luminosity increase in tandem (compared to the initial decline) while narrow optical emission lines corresponding to He\,\textsc{i} and both permitted and forbidden transitions of O\,\textsc{iii} appear in the spectrum. Given the high ionisation potential (higher than 35 eV), these nebular lines O\,\textsc{iii} lines typically only appear in the presence of a strongly ionising radiation field and relatively low densities. The absence of these lines in the early phases of the flare suggest that the ionising source was much fainter at those times. A scenario where we are witnessing the formation of an accretion disk during the {\it Swift} observations can explain the nebular lines if the disk radiation ionises debris (most likely the bound material, as the lines are observed at their rest wavelengths) from the disrupted star. The plateau in the lightcurve can then be explained as reprocessing of X-ray radiation into UV/optical photons, creating the right conditions for line emission. 
The disappearance of the nebular lines after the plateau indicate that the emitting layer of material has become fully ionised and optically thin to the X-ray radiation, which can explain the up-turn in the XRT lightcurve while the UV/optical emission becomes fainter. 

\subsection{Radio upper limits}
We can use the radio non-detections to constrain the presence of a jet/outflow similar to that observed in ASASSN--14li \citep{vanvelzen2015,Alexander2016,Romero2016,Pasham2018}. To this end, we assume that the scaling relation between the radio and X-ray luminosity of a tentative jet/outflow is similar to that of ASASSN--14li, L$_r$ $\propto$ L$_{x}^{2.2}$ \citep{Pasham2018}.
From Table \ref{tab:radio} we see that the observations can marginally rule out that such a jet was produced. 

If the X-ray-radio jet coupling was similar to that seen in ASASSN--14li, the difference in jet power could be explained by either a difference in available accretion power for the jet to tap into (assuming a similar jet efficiency), or by a difference in the conversion efficiency from accretion power to jet power \citep{Pasham2018}. While the latter is hard to test observationally, our observations disfavour the former scenario as the UV/optical and X-ray lightcurve and L$_{\rm opt}$/L$_{\rm x}$ evolution can potentially be explained by a relativistic encounter. \citet{Dai2015} have shown that this leads to higher accretion rates, hence this would result in a more powerful jet and more luminous radio emission if the jet power follows the mass accretion rate. 

One scenario that could explain the radio non-detection is the presence of a tenuous circumnuclear medium (CNM; \citealt{Generozov2017}). Unfortunately, for AT~2018fyk, no strong constraints can be made. This illustrates the need for deeper radio observations to rule out the presence of a jet, even in the case of a low density CNM. Upper limits several orders of magnitude deeper than those presented here are required to rule out a jet power similar to ASASSN--14li in known TDEs.

\begin{table}
\centering
\caption{Observed radio upper limits (stacked 16.7 and 21.2 GHz), compared to the radio luminosity expected for a radio - X-ray correlation similar to ASASSN--14li. The epoch denotes days after discovery.}
\begin{tabular}{ccc}\hline
Epoch & L$_{\rm radio}$ & L$_{\rm radio}$$\propto$L$_{x}^{2.2}$\\
(days) & (erg s$^{-1}$) & (erg s$^{-1}$) \\\hline\hline
11 & $<$ 5$\times$10$^{37}$ & 1$\times$10$^{36}$\\
38 & $<$ 1$\times$10$^{38}$ & 1$\times$10$^{38}$\\
75 & $<$ 8$\times$10$^{37}$ & 1$\times$10$^{38}$\\
\hline
\end{tabular}
\label{tab:radio}
\end{table}

\section{Summary}
\label{sec:summary}
We have presented and analysed multi-wavelength photometric and spectroscopic observations of the UV/optical and X-ray bright tidal disruption event AT~2018fyk. Gaia observations of the transient constrain the transient position to within $\sim$120 pc of the galaxy nucleus. The densely sampled {\it Swift} UVOT and XRT lightcurves show a peculiar evolution when compared to other well established TDEs but similar to ASASSN--15lh, including a secondary maximum after initial decline from peak. Optical spectra similarly showed peculiar features not previously identified, including both high and low ionisation narrow emission features. We show that similar features were present in archival spectra of at least one other TDE (ASASSN--15oi), but remained unidentified due to the complex line profiles of the broad emission lines. The main results from our analysis can be summarised as follows:

\begin{itemize}
\item The X-ray and UV/optical lightcurves show a plateau phase of $\sim$50 days after an initial monotonic decline. When the UV/optical decline resumes, the X-rays instead turn over and increase in luminosity. Such a two component lightcurve is similar to that seen in ASASSN--15lh, albeit on shorter timescales. It can be naturally explained in the scenario of a TDE with relativistic pericenter, where the disk formation process is fast and efficient, resulting in this second maximum to occur 10s-100s of days rather than 1000s of days after disruption, as observed for most TDEs \citep{vanvelzen2019}.\\

\item A high black hole mass ($\gtrsim 10^7$ M$_{\rm BH}$) can result in relativistic pericenters for a typical lower main sequence star. We therefore suggest that, similar to ASASSN--15lh, the peculiar lightcurve of AT2018fyk is due to the high M$_{\rm BH}$, which can provide the right conditions to explain the lightcurve shape. Moreover, we tentatively identify another double-peaked structure in the optical lightcurves of D3-13, which has M$_{\rm BH} \sim 10^{7.4}$ M$_{\odot}$. Double-peaked lightcurves might be a universal feature of TDEs around massive black holes (M$_{\rm BH} \gtrsim 10^7$ M$_{\odot}$) as the encounters are always expected to be relativistic.\\

\item The X-ray spectra can be relatively well described by an absorbed power-law + blackbody model (power-law index $\sim$3, kT$\sim$110 eV). The power-law contributes roughly 30 \% of the flux even at early times. In the final two epochs of observations before the source became Sun constrained, the spectrum appears to develop a harder component above 2 keV. Continued monitoring and analysis will reveal whether a hard power-law tail appears, or whether the spectrum remains dominated by the soft (blackbody) component.\\

\item The optical spectra show broad H$\alpha$ and \hetwo 4686 lines. We also tentatively identify broad Fe\,\textsc{ii} lines at 3449\AA\ and 3499 \AA. In particular the \hetwo 4686 line has a Gaussian FWHM significantly greater ($\sim$28$\times$10$^{3}$ km s$^{-1}$) than the other broad lines ($\sim$10-15$\times$10$^{3}$ km s$^{-1}$), suggesting it is a blend of multiple emission features.\\

\item We detect both high ionisation (O\,\textsc{iii}) and low ionisation (Fe\,\textsc{ii}) narrow emission lines. In particular the Fe\,\textsc{ii} complex near 4570 \AA\ is unambiguously detected. We propose that this line complex can explain the asymmetric line profiles in this and several other Fe-rich TDEs (e.g. ASASSN--15oi, PTF--09ge).\\

\item The presence of low ionisation Fe\,\textsc{ii} emission lines requires optically thick, high density gas and (most likely) a strong source of ionising photons. Taken together with the lightcurve evolution, this suggests that the X-ray radiation is (partially) being absorbed and efficiently re-emitted in the UV/optical. When the gas is sufficiently ionised it becomes optically thin to the X-rays, leading to a decline in the UV/optical emission and the observed increase in X-ray luminosity.\\

\item The spectral features are remarkably similar to those seen in NLS1 AGN, as well as very similar to other TDE candidates in AGN such as the extreme coronal line emitters. This suggests a connection between all these events around AGN and AT2018fyk, which occurred in a quiescent SMBH. This strengthens the arguments in favour of a TDE interpretation for PS16dtm, the \citet{Kankare2017} events and the coronal line emitters.\\
\end{itemize}

We have illustrated that a wealth of information can be extracted from contemporaneous X-ray and UV/optical observations made possible by {\it Swift} and spectroscopic monitoring, and shown the importance of dense temporal coverage to map the detailed behaviour of both the X-ray and UV/optical emission in TDEs. Increasing the sample of TDEs with such coverage will almost certainly lead to the discovery of new behaviour in these enigmatic cosmic lighthouses, which in turn will reveal the detailed physics that occurs in these extreme environments. The detection of narrow emission lines highlights the need for medium/high resolution spectroscopic follow-up of TDEs to uncover the full diversity of their optical spectral appearance.
\section*{Acknowledgements}
We are grateful for constructive remarks and suggestions from the referee. We also thank Richard Saxton for sharing the {\it Swift} data of XMMSL1 J0740, and Suvi Gezari for sharing some of the data in Figure \ref{fig:loptlx}.
TW is funded in part by European Research Council grant 320360 and by European Commission grant 730980. GL was  supported by a research grant (19054) from VILLUM FONDEN. JCAM-J is the recipient of an Australian Research Council Future Fellowship (FT 140101082). PGJ and ZKR acknowledge support from European Research Council Consolidator Grant 647208. MG is supported by the Polish NCN MAESTRO grant 2014/14/A/ST9/00121. KM acknowledges support from STFC (ST/M005348/1) and from H2020 through an ERC Starting Grant (758638). MN acknowledges support from a Royal Astronomical Society Research Fellowship. FO acknowledges support of the H2020 Hemera program, grant agreement No 730970. Based on observations collected at the European Organisation for Astronomical Research in the Southern Hemisphere under ESO programme 199.D-0143. We acknowledge the use of public data from the Swift data archive.
The Australia Telescope Compact Array is part of the Australia Telescope National Facility which is funded by the Australian Government for operation as a National Facility managed by CSIRO.
This work has made use of data from the European Space Agency (ESA) mission {\it Gaia} (\url{https://www.cosmos.esa.int/gaia}), processed by the {\it Gaia} Data Processing and Analysis Consortium (DPAC, \url{https://www.cosmos.esa.int/web/gaia/dpac/consortium}). Funding for the DPAC has been provided by national institutions, in particular the institutions participating in the {\it Gaia} Multilateral Agreement. We also acknowledge the Gaia Photometric Science Alerts Team (\url{http://gsaweb.ast.cam.ac.uk/alerts}).

\bibliographystyle{mnras.bst}
\bibliography{bibliography.bib}

\appendix
\section{{\it Swift} UVOT observations}
\onecolumn
\begin{table*}
\caption{{\it Swift} UVOT host unsubtracted photometry, in Vega magnitudes, and the {\it Swift} XRT count rates for each observation ID. The conversion factor from count rate to flux used in this work is 4.41\,$\times$\,10$^{-11}$. We provide the mean MJD of the reference times in the UVOT bands. This table will be made available in machine-readable form.} \label{tab:uvotmeasurements}
\begin{tabular}{cccccccc}
    \centering
    MJD & $U$ & $B$ & $V$ & $UVW1$ & $UVM2$ & $UVW2$ & XRT \\ \hline
    days & mag & mag & mag & mag & mag & mag & counts s$^{-1}$ \\\hline\hline
    58383.7279&15.9$\pm$0.06&16.96$\pm$0.07&16.39$\pm$0.09&15.1$\pm$0.04&14.92$\pm$0.03&14.61$\pm$0.03&0.005$\pm$0.0018 \\
58389.9486&16.0$\pm$0.07&16.92$\pm$0.08&16.72$\pm$0.14&15.13$\pm$0.05&14.94$\pm$0.04&14.76$\pm$0.04&0.055$\pm$0.0092 \\
58393.1195&16.01$\pm$0.09&17.01$\pm$0.11&16.42$\pm$0.15&15.33$\pm$0.06&14.98$\pm$0.05&14.8$\pm$0.04&0.033$\pm$0.0092 \\
58395.7247&16.05$\pm$0.07&17.19$\pm$0.09&16.54$\pm$0.12&15.4$\pm$0.05&15.05$\pm$0.04&14.9$\pm$0.04&0.025$\pm$0.0031 \\
58396.2499&16.07$\pm$0.09&17.17$\pm$0.11&16.54$\pm$0.15&15.39$\pm$0.06&15.07$\pm$0.05&14.9$\pm$0.04&0.020$\pm$0.0075 \\
58397.9151&16.18$\pm$0.06&17.09$\pm$0.07&16.59$\pm$0.1&15.44$\pm$0.04&15.22$\pm$0.04&15.05$\pm$0.03&0.048$\pm$0.0054 \\
58398.979&16.28$\pm$0.09&17.15$\pm$0.1&16.66$\pm$0.14&15.51$\pm$0.06&15.22$\pm$0.05&15.05$\pm$0.04&0.046$\pm$0.0091 \\
58399.7422&16.29$\pm$0.09&17.4$\pm$0.11&16.73$\pm$0.14&15.58$\pm$0.06&15.29$\pm$0.04&15.13$\pm$0.04&0.032$\pm$0.0039 \\
58401.1307&16.2$\pm$0.08&17.25$\pm$0.1&16.55$\pm$0.15&15.58$\pm$0.06&15.35$\pm$0.05&15.12$\pm$0.04&0.030$\pm$0.0049 \\
58403.7488&16.39$\pm$0.11&17.27$\pm$0.12&16.7$\pm$0.16&15.63$\pm$0.07&15.35$\pm$0.05&15.2$\pm$0.05&0.038$\pm$0.0100 \\
58404.5755&16.49$\pm$0.08&17.42$\pm$0.09&16.63$\pm$0.11&15.63$\pm$0.05&15.31$\pm$0.07&15.28$\pm$0.04&0.051$\pm$0.0064 \\
58406.1341&16.54$\pm$0.1&17.23$\pm$0.1&16.74$\pm$0.15&15.71$\pm$0.06&15.45$\pm$0.05&15.42$\pm$0.05&0.035$\pm$0.0077 \\
58408.4551&16.27$\pm$0.1&16.99$\pm$0.11&16.64$\pm$0.16&15.61$\pm$0.07&15.39$\pm$0.05&15.23$\pm$0.05&0.035$\pm$0.0084 \\
58409.9907&16.39$\pm$0.07&17.29$\pm$0.07&16.5$\pm$0.09&15.74$\pm$0.05&15.46$\pm$0.04&15.25$\pm$0.04&0.026$\pm$0.0047 \\
58412.3747&16.1$\pm$0.1&17.4$\pm$0.14&16.68$\pm$0.17&15.75$\pm$0.07&15.46$\pm$0.05&15.35$\pm$0.05&0.041$\pm$0.0100 \\
58413.9073&16.31$\pm$0.06&17.32$\pm$0.07&16.73$\pm$0.09&15.63$\pm$0.04&15.39$\pm$0.03&15.21$\pm$0.03&0.034$\pm$0.0048 \\
58415.9641&16.25$\pm$0.09&17.08$\pm$0.09&16.55$\pm$0.13&15.71$\pm$0.06&15.38$\pm$0.05&15.24$\pm$0.04&0.041$\pm$0.0086 \\
58416.1691&16.3$\pm$0.1&17.13$\pm$0.11&16.45$\pm$0.14&15.66$\pm$0.07&15.44$\pm$0.05&15.18$\pm$0.05&0.028$\pm$0.0083 \\
58417.3639&16.38$\pm$0.11&17.03$\pm$0.11&16.79$\pm$0.18&15.59$\pm$0.07&15.48$\pm$0.06&15.31$\pm$0.05&0.024$\pm$0.0079 \\
58417.8613&16.25$\pm$0.07&17.24$\pm$0.08&16.7$\pm$0.11&15.73$\pm$0.05&15.4$\pm$0.04&15.2$\pm$0.04&0.062$\pm$0.0061 \\
58418.6338&16.23$\pm$0.1&17.21$\pm$0.12&16.65$\pm$0.16&15.65$\pm$0.07&15.44$\pm$0.05&15.24$\pm$0.05&0.043$\pm$0.0098 \\
58420.1548&16.3$\pm$0.1&17.24$\pm$0.11&16.83$\pm$0.17&15.58$\pm$0.07&15.43$\pm$0.05&15.15$\pm$0.05&0.058$\pm$0.012 \\
58421.8815&16.29$\pm$0.07&17.44$\pm$0.09&16.47$\pm$0.1&15.61$\pm$0.05&15.45$\pm$0.04&15.22$\pm$0.04&0.033$\pm$0.0057 \\
58423.5402&16.31$\pm$0.09&17.13$\pm$0.1&16.42$\pm$0.12&15.7$\pm$0.06&15.38$\pm$0.05&15.16$\pm$0.04&0.035$\pm$0.0083 \\
58425.0384&16.24$\pm$0.06&17.18$\pm$0.07&16.59$\pm$0.1&15.58$\pm$0.04&15.34$\pm$0.04&15.2$\pm$0.03&0.066$\pm$0.0063 \\
58426.5273&16.22$\pm$0.09&17.23$\pm$0.11&16.68$\pm$0.16&15.54$\pm$0.07&15.36$\pm$0.06&15.19$\pm$0.05&0.053$\pm$0.012 \\
58427.323&16.3$\pm$0.1&17.08$\pm$0.1&16.49$\pm$0.14&15.57$\pm$0.06&15.34$\pm$0.05&15.17$\pm$0.05&0.035$\pm$0.0091 \\
58428.8499&16.28$\pm$0.09&17.29$\pm$0.11&16.52$\pm$0.13&15.69$\pm$0.06&15.3$\pm$0.04&15.12$\pm$0.04&0.088$\pm$0.012 \\
58429.8529&16.21$\pm$0.08&17.22$\pm$0.09&16.47$\pm$0.11&15.47$\pm$0.05&15.32$\pm$0.04&15.13$\pm$0.04&0.033$\pm$0.0075 \\
58430.0486&16.25$\pm$0.08&16.94$\pm$0.09&16.51$\pm$0.13&15.53$\pm$0.06&15.35$\pm$0.05&15.12$\pm$0.04&0.061$\pm$0.011 \\
58431.5081&16.21$\pm$0.09&17.14$\pm$0.1&16.64$\pm$0.15&15.58$\pm$0.06&15.35$\pm$0.05&15.1$\pm$0.04&0.071$\pm$0.011 \\
58434.9552&16.19$\pm$0.08&17.07$\pm$0.09&16.61$\pm$0.12&15.59$\pm$0.06&15.28$\pm$0.06&15.19$\pm$0.04&0.069$\pm$0.012 \\
58439.075&16.16$\pm$0.09&16.93$\pm$0.09&16.72$\pm$0.15&15.61$\pm$0.07&15.35$\pm$0.07&15.13$\pm$0.05&0.063$\pm$0.013 \\
58440.0052&16.31$\pm$0.08&17.24$\pm$0.09&16.72$\pm$0.12&15.44$\pm$0.06&15.35$\pm$0.05&15.16$\pm$0.04&0.033$\pm$0.0060 \\
58443.2648&16.33$\pm$0.08&17.18$\pm$0.08&16.62$\pm$0.12&15.63$\pm$0.06&15.35$\pm$0.06&15.19$\pm$0.04&0.039$\pm$0.0091 \\
58445.3327&16.28$\pm$0.07&17.2$\pm$0.08&16.74$\pm$0.12&15.63$\pm$0.06&15.43$\pm$0.06&15.31$\pm$0.04&0.024$\pm$0.0073 \\
58447.9741&16.46$\pm$0.09&17.2$\pm$0.1&16.65$\pm$0.13&15.74$\pm$0.07&15.41$\pm$0.08&15.26$\pm$0.04&0.049$\pm$0.011 \\
58449.0333&16.4$\pm$0.1&17.28$\pm$0.12&16.36$\pm$0.13&15.89$\pm$0.09&15.53$\pm$0.07&15.36$\pm$0.05&0.031$\pm$0.0090 \\
58451.3648&16.58$\pm$0.09&17.32$\pm$0.09&16.7$\pm$0.12&15.93$\pm$0.07&15.67$\pm$0.06&15.5$\pm$0.05&0.029$\pm$0.0074 \\
58455.5525&16.4$\pm$0.08&17.29$\pm$0.09&16.73$\pm$0.13&15.94$\pm$0.07&15.68$\pm$0.07&15.61$\pm$0.05&0.0069$\pm$0.0040 \\
58459.9956&16.8$\pm$0.11&17.4$\pm$0.11&16.6$\pm$0.12&16.24$\pm$0.09&15.88$\pm$0.07&15.64$\pm$0.05&0.016$\pm$0.0057 \\
58464.6536&16.63$\pm$0.07&17.47$\pm$0.08&16.68$\pm$0.1&16.07$\pm$0.06&15.9$\pm$0.06&15.76$\pm$0.04&0.021$\pm$0.0051 \\
58467.7725&16.69$\pm$0.07&17.51$\pm$0.08&16.68$\pm$0.09&16.2$\pm$0.06&15.96$\pm$0.06&15.78$\pm$0.04&0.020$\pm$0.0045 \\
58470.4318&16.81$\pm$0.08&17.52$\pm$0.08&16.79$\pm$0.1&16.31$\pm$0.07&16.07$\pm$0.06&15.84$\pm$0.04&0.022$\pm$0.0050 \\
58473.8851&16.86$\pm$0.08&17.47$\pm$0.08&16.7$\pm$0.09&16.13$\pm$0.06&15.92$\pm$0.06&15.82$\pm$0.04&0.029$\pm$0.0055 \\
58476.373&17.17$\pm$0.11&17.4$\pm$0.09&16.73$\pm$0.11&16.39$\pm$0.08&16.0$\pm$0.06&15.93$\pm$0.05&0.041$\pm$0.0045 \\
58479.9228&16.75$\pm$0.09&17.5$\pm$0.11&16.63$\pm$0.11&16.36$\pm$0.08&16.12$\pm$0.07&15.8$\pm$0.05&0.039$\pm$0.0076 \\
58482.0515&16.98$\pm$0.09&17.49$\pm$0.09&16.83$\pm$0.11&16.56$\pm$0.08&16.37$\pm$0.07&16.22$\pm$0.05&0.024$\pm$0.0055 \\
58485.2343&17.13$\pm$0.11&17.75$\pm$0.12&16.93$\pm$0.12&16.73$\pm$0.09&16.49$\pm$0.08&16.25$\pm$0.05&0.063$\pm$0.0089 \\
58491.9408&17.09$\pm$0.11&17.76$\pm$0.12&16.85$\pm$0.12&16.81$\pm$0.1&16.53$\pm$0.08&16.32$\pm$0.06&0.080$\pm$0.010 \\

    \hline
    \end{tabular}
\end{table*}
\label{lastpage}
\end{document}